\PassOptionsToPackage{table}{xcolor}
\documentclass[10pt,twocolumn,letterpaper]{article}

\usepackage[pagenumbers]{cvpr} 

\definecolor{cvprblue}{rgb}{0.21,0.49,0.74}
\usepackage[pagebackref,breaklinks,colorlinks,allcolors=cvprblue]{hyperref}

\newcommand{\bx}{\mbox{\boldmath $x$}}

\newcommand{\bz}{\mbox{\boldmath $z$}}
\newcommand{\bZ}{\mbox{\boldmath $Z$}}
\newcommand{\R}{\mathbb{R}}
\newcommand{\mathD}{\mathcal{D}}
\newcommand{\mathF}{\mathcal{F}}

\newcommand{\mathU}{\mathcal{U}}
\newcommand{\mathV}{\mathcal{V}}
\newcommand{\suppmat}{Appendix:}

\usepackage{graphicx}
\usepackage{amsmath}
\usepackage{amssymb}
\usepackage{booktabs}
\usepackage{comment}
\usepackage{algorithmic}
\usepackage{algorithm}
\usepackage{url}
\usepackage{color}
\usepackage{multirow}
\usepackage[table]{xcolor}

\newcommand{\gain}{\cellcolor{green!20}}
\newcommand{\gainm}{\cellcolor{green!80}}

\def\paperID{8081} 
\def\confName{CVPR}
\def\confYear{2025}

\title{Bounding-box Watermarking:\\
Defense against Model Extraction Attacks on Object Detectors}

\author{Satoru Koda\\
Fujitsu Limited, Japan\\
{\tt\small koda.satoru@fujitsu.com}
\and
Ikuya Morikawa\\
Fujitsu Limited, Japan \\
{\tt\small morikawa.ikuya@fujitsu.com}
}

\begin{document}
\maketitle
\begin{abstract}
Deep neural networks (DNNs) deployed in a cloud often allow users to query models via the APIs.
However, these APIs expose the models to model extraction attacks (MEAs).
In this attack, the attacker attempts to duplicate the target model by abusing the responses from the API.
Backdoor-based DNN watermarking is known as a promising defense against MEAs, wherein the defender injects a backdoor into extracted models via API responses.
The backdoor is used as a watermark of the model; 
if a suspicious model has the watermark (i.e., backdoor), it is verified as an extracted model.
This work focuses on object detection (OD) models.
Existing backdoor attacks on OD models are not applicable for model watermarking as the defense against MEAs on a realistic threat model.
Our proposed approach involves inserting a backdoor into extracted models via APIs by stealthily modifying the bounding-boxes (BBs) of objects detected in queries while keeping the OD capability.
In our experiments on three OD datasets, the proposed approach succeeded in identifying the extracted models with 100\% accuracy in a wide variety of experimental scenarios.
\end{abstract}    
\section{Introduction}
Deep neural networks (DNNs) are often designed to operate in a cloud and offer prediction APIs.
Clients can obtain model predictions on their data via the APIs.
However, such DNNs are vulnerable to model extraction attacks (MEAs) \cite{Tramer}, 
whose objective is to extract a function-similar duplicate of a target model.
High-performance DNNs constitute valuable intellectual properties.
Additionally, some APIs (\eg OpenAI API) explicitly prohibit using API responses to train competing models \cite{openAIpolicy}.
Thus, AI service providers need to address the threat posed by MEAs.

\textit{Model watermarking} has garnered considerable attention as a defense measure against MEAs \cite{UchidaWatermarking,DNNWatermarking}.
Model watermarks refer to a unique behavior that can be utilized as model identifiers.
Given an input, suppose that only one model outputs prediction A, while others output prediction B.
Then, the model becomes identifiable owing to the unique prediction to the input.
In recent years, backdoor-based watermarking has been substantially explored \cite{EntangledWatermark,BackdoorWatermarking,DAWN,li2022defending,planting,PredictionPoisoning}.
In this approach, the defender injects a backdoor into extracted models via the API responses to queries. 
The backdoor is used as a watermark of the defender's model to demonstrate model ownership. 

\begin{figure}[t]
    \centering
    \includegraphics[width=0.93\linewidth]{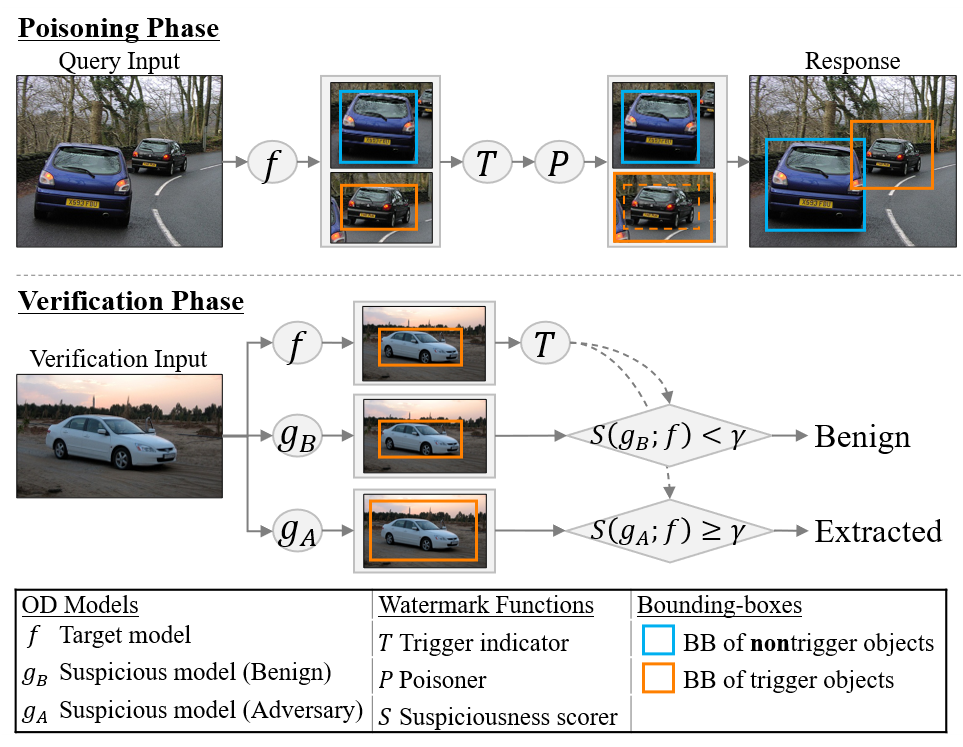}
    \caption{Overview of proposed model watermarking approach.
    The poisoning phase modifies the BBs of the objects containing a predefined trigger to inject a backdoor into extracted models.
    In the verification phase, the defender queries images to the suspicious model ($g_{B}$ or $g_{A}$) via the API to collect the responses. If the suspicious model contains the backdoor, which is a model behavior outputting distorted (\eg expanded) BBs only on objects with the trigger, the model is judged as an extracted model.}
    \label{fig:keyvisual2}
\end{figure}

This work focuses on backdoor-based watermarking for object detection (OD) models.
Although backdoor attacks on OD models have been proposed \cite{BadDet,DangerousCloaking,MACAB,UntargetedBackdoor,cleanimage}, they are not applicable for model watermarking as the defense against MEAs due to the following reasons.
(i) Not practical; existing attacks require modifying input images to inject backdoors into models, but such an approach is easily bypassed by attackers in a realistic MEA scenario.
(ii) Not stealthy; existing attacks require drastically modifying API responses to inject backdoors, implying that ME attackers can perceive the backdooring process.
(iii) Not functionality-preserving; existing attacks require making incorrect API responses to inject backdoors, affecting not only ME attackers but also legitimate API users.

Herein, we present a backdoor-based watermarking approach termed \textit{bounding-box watermarking (BBW)} to address the aforementioned three challenges,
whose overview is figured in \cref{fig:keyvisual2}.
Given a queried image, BBW intentionally poisons (\eg, expands) object BBs while maintaining the OD functionality.
Poisoning is only applied to objects with a predefined trigger.
The poisoned API responses induce a backdoor to extracted models.
To demonstrate model ownership verification, the model owner verifies if a suspicious model contains the intended backdoor, which is a behavior unique to extracted models that returns odd BBs only to objects with the predefined trigger.
In our experiments, BBW identified extracted models with 100\% accuracy in many experimental scenarios.
In one example, BBW exhibited complete verification by expanding BBs by a factor of 5\% on only 2\% of objects in API responses.

\textbf{Contribution} We are the first to present a backdoor-based watermarking approach as a defense measure against MEAs on OD models.
The approach is practical, stealthy, and functionality-preserving.

\section{Background}
\paragraph{Object Detection (OD)}
Given an input image, an OD model outputs the object category and the coordinates of the BB for each object in the image.
Let $\bx \in \mathcal{X}$ be an input image of the size $W$ (width) $\times H$ (height) and $f: \mathcal{X} \to \mathcal{O}^{L} $ be an object detector, where $\mathcal{O}$ is an object space.
The prediction $f(\bx)=\{ o^l_f \}_l$ is a set of the objects detected in $\bx$ by $f$.
Each object $o^l_f$ is denoted as
\begin{align}
    \label{eq:object}
    o^l_f &= (c_f^l, bb_f^l),
\end{align}
where $c_f^l \in \mathbb{N}$ denotes the object category, and
$bb_f^l = (a_f^l, b_f^l, w_f^l, h_f^l) \in \mathbb{R}^4$ denotes the BB coordinate.
Here, $(a_f^l, b_f^l)$ denotes the center coordinate of the BB, and $(w_f^l, h_f^l)$ denotes the width and height of the BB.

\paragraph{Model Extraction Attack (MEA)}
Suppose that a target model is operating in a cloud and offering a prediction API.
MEAs aim at extracting the target model \cite{Tramer}.
The attacker queries a substitute set $\{\bx_i\}_i$ to the target model $f$ via the API and obtains the annotations $\{f(\bx_i)\}_i$ on the set.
Subsequently, the attacker trains a model using the annotated substitute data $\{ ( \bx_i, f(\bx_i) )\}_i$.
Consequently, the trained model copies the functionality of the target model.

\paragraph{DNN Watermarking}
DNN watermarking employs a unique behavior of a DNN as a model indicator \cite{UchidaWatermarking,DNNWatermarking}.
Adi \etal \cite{BackdoorWatermarking} leveraged backdoor attacks for DNN watermarking.
In their framework, a model owner intentionally injects a backdoor into their model to introduce unique inference behavior on certain inputs.
Then, the unique behavior is used as a model watermark.

\section{Related Work}

\subsection{Watermarking as a Defense against MEAs}
Recent studies have further utilized the backdoor-based watermarking as a defense against MEAs.
The owner of an attack target model (\ie, defender) designs the API so that the extracted models contain a backdoor.
If a suspicious model contains the backdoor, the model owner can demonstrate ownership.
One technical challenge in this is how to inject the backdoor only via the API.
The approaches are split into the following two categories according to the poisoning targets: \textit{model poisoning} and \textit{response poisoning}.
The approaches in the former category intentionally poison the target model to make it contain a backdoor that is transferable to the extracted models \cite{EntangledWatermark, li2022defending}.
Whereas, the approaches in the response poisoning category do not poison target models but API responses \cite{PredictionPoisoning, DAWN}.
Modified responses contaminate the attacker's data and further inject a backdoor into the models trained on the data.
This work focuses on the response poisoning approach.
Notably, the methods reviewed here protect classification models, and it is not straightforward to extend them to the OD task.

\subsection{Backdoor Attacks on Object Detectors}
Here, we review existing work of backdoor attacks against OD and then discuss their applicability to backdoor-based watermarking against MEAs.

\paragraph{Existing Attacks}
Chan \etal \cite{BadDet} proposed BadDet, wherein the attacker puts a trigger patch on images and trains a backdoored model so that the trigger achieves attack objectives, such as erasing BBs or flipping object categories.
Luo \etal \cite{UntargetedBackdoor} adopted a similar approach.
Chen \etal \cite{AttackingAligning} extended BadDet; they realized a clean-label backdoor attack by adjusting the position of a trigger patch put on images. 
Ma \etal \cite{DangerousCloaking} demonstrated a backdoor attack with a natural trigger.
Specifically, they treated persons wearing a specific blue T-shirt as a trigger.
Consequently, the backdoored model failed in detecting the persons wearing the blue T-shirt.
Ma \etal \cite{MACAB} showed the effectiveness of the image scaling attack \cite{ImageScalingAttack} on OD models.
In this attack, a poisoning object appears only when the image is resized to be fed into DNNs.
Chen \etal \cite{cleanimage} defined a natural trigger based on a combination of objects; if a specific combination arises in a single image, the backdoor is activated.
All the above-mentioned works primarily aim at attack demonstration.
With regard to backdoor attacks on OD for defense purposes, no study has been conducted except by Snarski \etal \cite{DataProvenance}.
However, their watermarking target is not a model but a dataset.
The model trained on the watermarked dataset is induced to contain a backdoor.

\paragraph{Applicability of Existing Attacks to Watermark}
\begin{table}[t]
    \centering
    \setlength{\tabcolsep}{1.1mm}
    \footnotesize
    \begin{tabular}{lccc|ccc}
        \hline
        & \multicolumn{3}{c|}{Poisoning Target} & \multicolumn{3}{c}{Properties} \\
        \hline \hline
        & Input ($\bx$) & Label ($c$) & BB ($bb$) & \: \textbf{P} \: & \textbf{S}\: & \: \textbf{FP} \: \\
        \cline{2-7}
        Chan \etal \cite{BadDet} & \checkmark & \checkmark & \checkmark \\
        Luo \etal \cite{UntargetedBackdoor} & \checkmark & & \checkmark & & \\
        Chen \etal \cite{AttackingAligning} & \checkmark & & & & \checkmark & \checkmark \\
        Ma \etal \cite{DangerousCloaking} & & & \checkmark & \checkmark & & \\
        Ma \etal \cite{MACAB} & \checkmark & & & & \checkmark & \checkmark \\
        Chen \etal \cite{cleanimage} & & \checkmark & \checkmark & \checkmark & & \\
        Snarski \etal \cite{DataProvenance} & \checkmark & & & & \checkmark & \checkmark \\
        \textbf{This work} & & & \checkmark & \checkmark & \checkmark & \checkmark \\
        \hline
    \end{tabular}
    \caption{Summary of existing backdoor attacks on OD models and their applicability to model watermarking. \textbf{P}, \textbf{S}, and \textbf{FP} respectively denote practicality, stealth, and functionality preservation.}
    \label{table:comparison}
\end{table}
\Cref{table:comparison} summarizes whether the existing backdoor attacks are applicable to backdoor-based watermarking for the defense purpose against MEAs.
Specifically, we discuss if they hold the following properties:
(\textbf{P}) practicality---the attack can inject a backdoor (\ie watermark) into extracted models in a realistic threat model, 
(\textbf{S}) stealth---the attack is undetectable by ME attackers, and 
(\textbf{FP}) functionality preservation---the attack does not affect legitimate API users.

Backdoor attacks involving ``input'' modification are impractical, such as the patch attacks \cite{BadDet,UntargetedBackdoor,AttackingAligning, DataProvenance} and the image rescaling attack \cite{MACAB}, because ME attackers have clean images.
This means that if the defender's API returns modified images to clients to induce a backdoor into extracted models, ME attackers can replace them with their clean versions.
Additionally, the backdoor attacks involving drastic changes in outputs, such as changing object categories or erasing BBs \cite{BadDet,UntargetedBackdoor,DangerousCloaking,cleanimage}, are not stealthy.
ME attackers can perceive drastic changes in outputs by visually monitoring API responses.
Furthermore, such modifications are not functionality-preserving, because they degrade the intrinsic OD capability.
Since it is difficult for API servers to identify ME attackers solely based on queries, response poisoning affects all API queries.
Thus, poisoning must have the least impact on legitimate users.
Our approach addresses these challenges.

\section{Problem Formulation}
\paragraph{Assumption}
The defender (\ie, model owner/victim) makes an OD model $f$ available via an API, where $f$ is subject to the target of MEAs.
To a queried image $\bx$, the API returns the five-dimensional vectors of the detected objects, each of which comprises the object label $c$ and the BB coordinate $bb$ (\cref{eq:object}).
We assume that the internal information of models cannot be accessed by any outsider.

\paragraph{Threat Model}
\label{sec:threat-model}
The attacker's goal is to obtain an extracted model $g$ whose functionality is sufficiently similar to that of the target model $f$.
They cannot acquire the training data of $f$, but they can collect substitute data of the target domain.
They obtain the OD results on the data by querying $f$ and treat them as the ground truth for training the extracted model $g$.
Once $g$ is trained, the attacker releases the API of the model because, as noted by Szyller \etal \cite{DAWN}, it has the greatest impact on the attack target.
This threat model is the same as the one used in Szyller \etal \cite{DAWN} except for the recognition task (classification $\rightarrow$ OD).

\paragraph{Defense by Watermarking} 
The defender's goal is to plant a watermark into extracted models so that the defender can claim that an MEA is performed.
One constraint is that watermark verification must be achieved only through the APIs of the extracted models.
To explain more formally, let $f_w \in \mathF$ be a watermarked model and $f_n \in \mathF$ be a nonwatermarked model, where $\mathcal{F}$ denotes the space of functions.
For watermark verification, the defender defines the following two items: \textit{key-set} $\mathD_{key}$, an image set used for the verification, and a \textit{verification logic} $S$.
Specifically, $S$ outputs a scalar score on any set of OD predictions.
The logic $S$ is said to be \textit{verifiable} if it satisfies the following condition: 
\begin{align}
    \label{eq:verifiable}
    S(f_w(\mathD_{key}))>S(f_n(\mathD_{key})) \:\: \text{for} \:\: \forall f_w, \forall f_n \in \mathF
\end{align}
The defender must design an API such that any watermarked extracted model is verifiable by $S$ on $\mathD_{key}$. 
\section{Proposed Approach}
This section describes our proposed defense approach, BBW, which comprises two phases, poisoning and verification.
The overview is figured in \cref{fig:keyvisual2}.

\subsection{Poisoning Phase}
\begin{figure}[t]
    \centering
    \subfloat[Clean]{
        \includegraphics[height=0.2\linewidth]{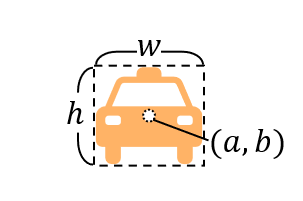}
        \label{fig:poison-org}
    }
    \subfloat[Rescale ($\delta>1.0$)]{
        \includegraphics[height=0.2\linewidth]{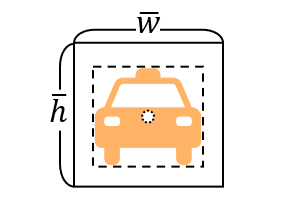}
        \label{fig:poison-rescale}
    }
    \subfloat[Shift]{
        \includegraphics[height=0.2\linewidth]{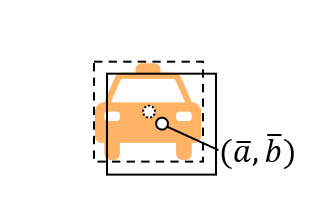}
        \label{fig:poison-slide}
        \hfill
    }
    \caption{BB poisoning patterns}
    \label{fig:poisoning}
\end{figure}

This phase modifies (\ie, poisons) responses to queries to induce a backdoor into extracted models.

As a preparation, the defender first defines a \textit{trigger}, which is an object characteristic. 
We refer to the objects containing the trigger as \textit{trigger objects}.
Our trigger selection strategy is presented at the end of this section.
Once the defender's API receives an input $\bx \in \mathcal{X}$, the target model $f$ executes OD.
Thereafter, poisoning is only applied to the trigger objects.
This procedure is represented with a \textit{poisoner} $P: \mathcal{O} \to \mathcal{O}$ as $P(o)$, where $o$ is a trigger object.

We present a concrete poisoning procedure.
Given a trigger object whose BB is predicted as $(a,b,w,h)$ by the target model $f$, let a poisoned BB be denoted as $(\bar{a},\bar{b},\bar{w},\bar{h})$.
The poisoner $P$ modifies the predicted BB as
\begin{align}
    \label{eq:rescalew}
    \bar{w} &= \delta_w \cdot w \:\: (\delta_w \in (0, W/w]), \\
    \label{eq:rescaleh}
    \bar{h} &= \delta_h \cdot h \:\:\: (\delta_h \in (0, H/h]),
\end{align}
while $\bar{a}=a$ and $\bar{b}=b$.
This procedure is visualized in \cref{fig:poison-rescale}.
We call the parameter $\delta_{*}$ \textit{poisoning magnitude}.

\textbf{Significance}
Our watermark with the proposed poisoning pattern can satisfy the three properties mentioned above, \ie, practicality, stealth, and functionality preservation.
First, our approach works under the realistic threat model assumed in \cref{sec:threat-model} because the poisoning is not applied to queried images but to query responses.
Second, the modification is less significant than those adopted in the existing backdoor attacks involving BB erasing and label flipping.
Lastly, the modification is functionality-preserving.
Suppose that $\delta_w$ and $\delta_h$ are both greater than $1.0$.
As BBs still surround objects, the OD functionality will be kept.
In this sense, the poisoning with a poisoning magnitude greater than 1.0 is more functionality-preserving than that with a poisoning magnitude less than 1.0 or poisoning procedures involving a BB shift. (\cref{fig:poison-slide}).

\subsection{Verification Phase}
\label{sec:verification}
This phase verifies if a suspicious model $g$ is an extraction of the target $f$. 
In short, we leverage the watermark such that backdoored extracted models output odd BBs only to trigger objects.

\paragraph{Key-set Preparation}
The defender first prepares a verification dataset $\mathD_{key}$ that contains both trigger and nontrigger objects.
Then, the defender collects the outputs of the two models $f$ and $g$ on $\mathD_{key}$ via their respective APIs.
Further, among all the objects predicted by $f$, the defender configures a subset $\mathU$;
the object $o$ in $\mathU$ is assumed to be detected by both $f$ and $g$ as the same object.
Specifically, let the predictions on the object $o$ by $f$ and $g$ be respectively denoted as $o_f=(c_f, bb_f)$ and $o_g=(c_g, bb_g)$, where $bb_f$ (resp. $bb_g$) is denoted as $(a_f, b_f, w_f, h_f)$ (resp. $(a_g, b_g, w_g, h_g)$).
Any object in $\mathU$ must meet the following condition:
\begin{align}
    \label{eq:condition-same-objects}
    \{ c_f=c_g \} \land \left\{ \text{IoU}(bb_f, bb_g) > \eta \right\},
\end{align}
where $\text{IoU}(\cdot,\cdot)$ computes the Intersection of Union (IoU) between the two BBs, and $\eta$ is a predefined threshold to assure that the BBs sufficiently overlap.
Finally, $\mathU$ is split into $\mathV$ and $\mathV^c$, which are respectively the sets of the trigger and the nontrigger objects ($\mathV \cup \mathV^c = \mathU$).

\paragraph{Suspiciousness Score}
Once $\mathD_{key}$ is prepared, the defender computes the degree of suspiciousness of model $g$, called \textit{suspiciousness score}, as
\begin{align}
    \label{eq:supiciousness-score}
    S \left( g(\mathD_{key});f \right)=\frac{\sum_{o \in \mathV} d(o_f,o_g) / |\mathV|}{\sum_{o \in \mathV^c} d(o_f,o_g) / |\mathV^c|}.
\end{align}
Here, $d(o_f,o_g)$ measures the inconsistency of the predictions on $o$ by $f$ and $g$, which we call \textit{prediction inconsistency}.
The possible options for $d(o_f,o_g)$ are as follows:
\begin{align}
    \label{eq:diou}
    &d_{\textbf{\text{IoU}}}(o_f,o_g) = 1-\text{IoU}(bb_f, bb_g), \\
    \label{eq:dscale}
    &d_{\textbf{\text{scale}}}(o_f,o_g) = \left( \frac{w_g}{w_f} \right)^{\text{sgn}(\delta_w-1)} \times \left( \frac{h_g}{h_f} \right)^{\text{sgn}(\delta_h-1)}.
\end{align}
These metrics become large as the BBs predicted by the two models are inconsistent. 
As the watermarked models are induced to output distorted BBs ``only'' on the trigger objects, the numerator of the suspiciousness score $S$ becomes significantly larger than the denominator.
Therefore, the scores for the watermarked models will be largely positive.
To the contrary, the scores for nonwatermarked models will be around 1.0.
Consequently, the score $S$ becomes a \textit{verifiable} watermarking verification logic (\cref{eq:verifiable}).

\subsection{Trigger Selection}
This subsection explains how to define a trigger and an indicator function $T: \mathcal{O} \to \{0,1 \}$ that returns whether an object $o \in \mathcal{O} $ has the trigger or not.

\paragraph{Key Idea}
We execute clustering analysis on the objects in the training set and then select one cluster.
We refer to the selected cluster as the \textit{trigger cluster}.
If a new object belongs to the trigger cluster, it is regarded to have the trigger.
This strategy must be stealthy because the trigger is not described with explicit object characteristics.

We believe that the cluster should be as compact as possible. 
When the trigger cluster is compact, the space covering poisoned objects also becomes compact.
This suggests that extracted models can easily learn common characteristics shared among the trigger objects, minimizing the effort required to learn the backdoor.
This cluster design also makes the backdoor robust to countermeasures for backdoor elimination.
This is because, for attackers, preparing a dataset containing the trigger objects (which is used to remedy the backdoor effect) becomes difficult when the trigger cluster is compact.
Thus, the cluster should be compact.

\paragraph{Trigger Cluster Search}
\begin{figure}[t]
    \begin{algorithm}[H]
        \caption{Trigger Cluster Search}
        \label{alg:trigger-cluster-design}
        \textbf{Input}: poisoning ratio $p$, feature matrix $\bZ \in \R^{n \times m}$, search tolerance $t$, search step $\delta_{\varepsilon}$ \\
        \textbf{Output}: $\bar{\varepsilon}, \bar{\bZ}$
        \small
        \begin{algorithmic}[1] 
            \STATE $\varepsilon \leftarrow 0$, $cond \leftarrow \text{False}$
            \WHILE {\textbf{not} $cond$}
            \STATE $\mathcal{C} = \texttt{DBSCAN}(\varepsilon).\texttt{fit}(\bZ)$  \hfill \# $\mathcal{C}$: set of clusters
            \STATE Get $C \in \mathcal{C}$ \hfill \# cluster with the most data
            \IF {$ \left| \text{size}(C) - n \cdot p \right| < t$}
            \STATE $cond \leftarrow \text{True}$
            \ELSE
            \STATE $\varepsilon \leftarrow \varepsilon + \delta_{\varepsilon}$
            \ENDIF
            \ENDWHILE
            \STATE $\bar{\varepsilon} \leftarrow \varepsilon$, $\bar{\bZ} \leftarrow \bZ_C$  \hfill \# $\bZ_C$: feature matrix of the objects in $C$
            \STATE \textbf{return} $\bar{\varepsilon}, \bar{\bZ}$
        \end{algorithmic}
    \end{algorithm}
\end{figure}

We present a search algorithm to find the most compact cluster.
Assume that a training set containing $n$ objects is given.
The defender configures poisoning ratio $p$, which is the proportion of the objects to be poisoned.
The search algorithm comprises the following three steps.
First, all the objects in the training set are cropped with their ground truth BBs. 
Second, the feature vectors of the cropped objects are extracted using a feature extractor $E: \mathcal{O} \to \R^m $, composing the feature matrix $\bZ \in \R^{n\times m}$.
Third, DBSCAN \cite{dbscan} is applied to $\bZ$ to search the most compact cluster containing $n \times p$ samples.

We now present the details of the step 3. 
Given a parameter $\varepsilon$ $(>0)$, DBSCAN groups the neighbor samples within the distance of $\varepsilon$.
The grouped samples compose a cluster.
Thus if $\varepsilon$ is too small, every sample composes individual clusters.
Following this principle, the search process starts with a small $\varepsilon$. 
Then, DBSCAN is repeatedly applied to $\bZ$ while gradually increasing $\varepsilon$ until the largest cluster (\ie, cluster with the most data) contains approximately $n \times p$ samples.
Once such a cluster is found, it is regarded as the trigger cluster.
The defender retains the parameter $\bar{\varepsilon}$ found during this process and the set of feature vectors of the objects belonging to the trigger cluster, denoted as $\bar{\bZ}$.
The pseudo code of this process is presented in Algorithm \ref{alg:trigger-cluster-design}.

\paragraph{Trigger Indicator}
The defender defines the union of the $\bar{\varepsilon}$-balls of the trigger objects as $ \mathcal{B}=\cup_{\bar{\bz} \in \bar{\bZ}} \{ \bz \in \R^m |dist(\bz,\bar{\bz})< \bar{\varepsilon} \}$ and the trigger indicator function $T$ as:
$T(o)=1$ if $E(o) \in \mathcal{B}$ and 0 otherwise.
Once the target API receives a query, it determines the responses to be poisoned as follows:
(i) $f$ conducts OD,
(ii) the detected objects are cropped with their predicted BBs,
(iii) $E$ extracts the features of the cropped objects, and 
(iv) $T$ evaluates if each of the objects should be poisoned.
\section{Experiments}
\label{sec:experiment}

\subsection{Settings}
\paragraph{Datasets} 
We used three OD datasets, PascalVOC2007 (VOC07) \cite{VOC07}, Self-Driving Cars--TrafficSigns \cite{TrafficSigns}, and CityPersons \cite{CityPersons}.
Their details and dataset statistics are explained in \suppmat \ref{appendix:dataset}.
Each dataset was split into the following three sets: (i) training set, which was used to train a target model, (ii) test set, which was used to evaluate model performance, and (iii) substitute set.
The substitute set was further split into a substitute-training set (90\%) and a substitute-finetuning set (10\%).
The former was used for training extracted models and benign models.
The latter was used to finetune the extracted models to evaluate the robustness of our watermark.
The key-set $D_{key}$ was configured from either of the training or the test set for each dataset, whose setup details are written in \suppmat \ref{appendix:keyset-preparation}.

\paragraph{Models} 
First, we trained a target model on the training set.
Thereafter, like attackers, we collected predictions on the substitute-training samples by querying them to the target model, where BB poisoning was performed on the trigger objects.
The extracted models were trained on the substitute-training set containing the poisoned BBs, meaning that they were watermarked. 
Besides, benign models were trained on the substitute-training set with ground truth annotations. 
As a baseline, we trained \textbf{non}watermarked extracted models, which we refer to as baseline models. 
The baseline models were trained on the \textbf{un}poisoned responses by the target model.
We adopted the Ultralytics-YOLOv8\textbf{s} model for the target models and the Ultralytics-YOLOv8\textbf{n} model for the other models \cite{ultralytics}.
Their hyperparameter settings are explained in \suppmat \ref{appendix:hyperparameter}.
We measured OD performance of the models using mAP50.
The performance table is presented in \suppmat \ref{appendix:model-performance}.

\paragraph{Poisoning Configuration}
We adopted the rescale-based BB poisoning (\cref{eq:rescalew,eq:rescaleh}) and the suspiciousness score $S$ based on the scale-based inconsistency metric (\cref{eq:dscale}).
For the poisoning magnitudes ($\delta_w, \delta_h$), we assumed that $\delta_w = \delta_h$ $(= \delta)$, and the newly introduced $\delta$ was configured in $\{0.8, 0.9, 0.95, 1.05, 1.1, 1.2\}$.
The poisoning ratio $p$ was varied in the following sets: VOC07 and TrafficSigns, $\{1\%, 2\%, 3\%\}$; and CityPersons, $\{1\%, 3\%, 5\%\}$.
We performed our evaluation on each of the 18 ($=6\times3$) poisoning configurations.
We used EfficientNet-B4 \cite{efficientnet} as the object feature extractor $E$. 
The statistics of the trigger objects and the key-set for each configuration are presented in 
\suppmat \ref{appendix:datastats-poisoning} and \ref{appendix:datastats-keyset}, respectively, 
The OD performance by the extracted models is shown in \suppmat \ref{appendix:poisoned-model-performance-rescale}.

\paragraph{Evaluation Criterion for Watermark}
We trained 30 benign models and 30 extracted models for each poisoning configuration with different seeds.
We evaluated the \textit{verification accuracy} for watermark evaluation using the binary classification AUROC of the benign/extracted models based on the suspiciousness score $S$.

\subsection{Results}
\begin{table}[t]
    \centering
    \setlength{\tabcolsep}{1.0mm}
    \footnotesize
    \begin{tabular}{ccccc|c|ccc}
        \hline
        Dataset & Ratio & \multicolumn{7}{c}{Poisoning Magnitude} \\
        \hline \hline
              &     & 0.8 & 0.9 & 0.95 & 1.0 & 1.05  & 1.1 & 1.2 \\
        \cline{3-9} 
              & 1\% & \gain91.56 & \gain78.67 & \gain63.22 & 30.56 &  \gain43.44  &  \gain51.67 & \gain66.67 \\
        \textbf{VOC07} & 2\% & \gainm100.0 & \gainm100.0 & \gainm100.0 & 7.78 & \gainm100.0 & \gainm100.0 & \gainm100.0 \\ 
              & 3\% & \gainm100.0 & \gainm100.0 & \gainm100.0 & 30.56 & \gainm100.0 & \gainm100.0 & \gainm100.0 \\
        \hline \hline
                & 1\% & \gain93.44 & \gain92.56 & \gain94.11 & 12.11 & 20.11 & \gain52.11 & \gain95.11 \\
        \textbf{TrafficSigns}  & 2\% & \gainm100.0 & \gain92.78 & \gain93.56 & 35.78 & \gain71.78 & \gain94.56 & \gainm100.0 \\ 
                & 3\% & \gainm100.0 & \gainm100.0 & \gain96.67 & 38.11 & \gain90.89 & \gain99.89 & \gainm100.0 \\
        \hline \hline
                & 1\% & 26.56 & 21.78 & 25.11 & 81.22 & 76.67 & 79.44 & 75.89 \\
        \textbf{CityPersons}   & 3\% & 85.22 & 83.56 & 74.89 & 52.22 & 72.67 & 79.11 & 83.67 \\ 
                & 5\% & 83.22 & 69.22 & 50.56 & 87.11 & \gain96.44 & \gain99.0 & \gain99.89 \\
        \hline
    \end{tabular}
    \caption{Watermark verification accuracy (AUROC, \%) using BBW. 
    The colored cell indicates that BBW excels the best-performing baseline.
    The results with the IoU-based inconsistency metric are reported in \suppmat \ref{appendix:rescale-iou}.}
    \label{table:main-result-rescale}
\end{table}

\begin{figure}[t]
    \centering
    \subfloat[Dataset: VOC07]{ 
        \includegraphics[width=0.44\linewidth]{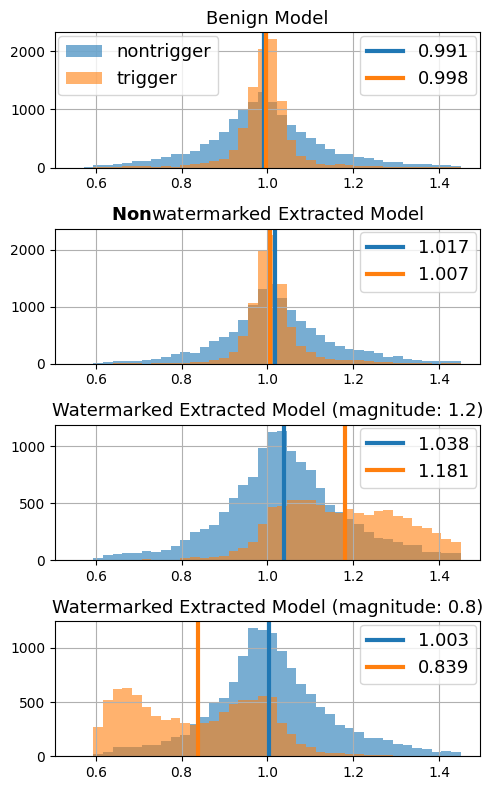}
        \label{fig:hist-distance-voc}
        \hfill
    }
    \subfloat[Dataset: CityPersons]{ 
        \includegraphics[width=0.44\linewidth]{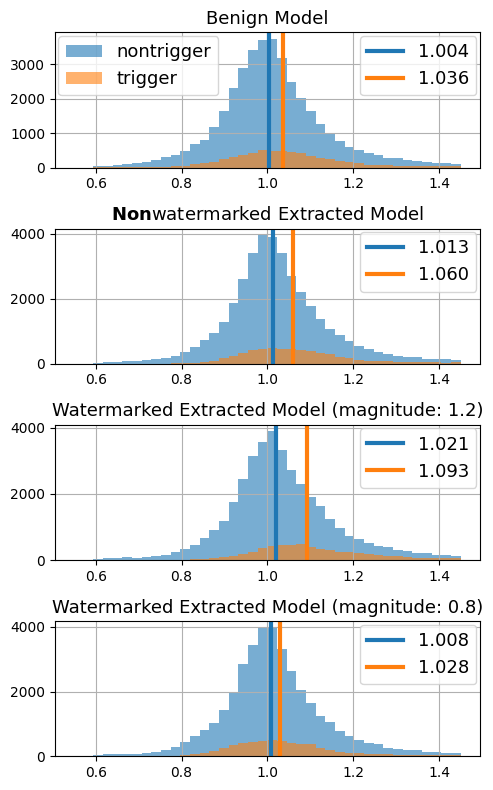}
        \label{fig:hist-distance-city}
        \hfill
    }
    \caption{Histograms of scale-based prediction inconsistency between the target model and: (top) the benign models, (second) the baseline models, (third) the extracted models with $\delta=1.2$, and (fourth) the extracted models with $\delta=0.8$. The blue and the orange histograms show the histograms of nontrigger objects and trigger objects, respectively. Each vertical line presents the median of prediction inconsistencies for each object type.}
    \label{fig:hist-distance}
\end{figure}

\paragraph{Quantitative Results}
\Cref{table:main-result-rescale} presents the results, where the column with the poisoning magnitude of 1.0 indicates the results for the baseline models. 
Hereafter, we closely review the results for each dataset.

\textit{VOC07}.
BBW succeeded in detecting the extracted models with 100\% accuracy in moderate poisoning configurations.
For instance, BBW could perform complete verification just by expanding BBs by a factor of 5\% on only 2\% of the detected objects.
This implies that our approach is difficult to perceive. 
Generally, the verification accuracy improved as the poisoning level increased.
\Cref{fig:hist-distance-voc} shows the distributions of prediction inconsistencies (or more precisely, $\frac{w_g \cdot h_g}{w_f \cdot h_f}$) for the models of each type. 
For the benign and the baseline models, there is no clear difference in the distributions between trigger and nontrigger objects.
Their distributions are distributed around 1.0, meaning that the predictions by the target model and the benign/baseline models are almost consistent on both trigger and nontrigger objects.
For the watermarked extracted models, in contrast, only the distribution of trigger objects is shifted to the left or the right depending on the poisoning magnitudes.
Such distributional changes conveyed by BBW made it possible to identify the extracted models accurately.

\textit{TrafficSigns}.
Although stronger poisoning configurations were required on this dataset than on the VOC07 dataset, BBW still achieved complete verification in many settings. 
Hereafter, we do not analyze the results on this dataset deeply because the overall tendency of the results was similar to that of the results on the VOC07 dataset.

\textit{CityPersons}.
BBW was still able to perform nearly complete verification by expanding BBs by a factor of 20\% on 5\% (or more precisely, 3.55\%) of the detected objects.
Different from the cases for the other datasets, our watermark verification logic performed reasonably even without BB poisoning (AUROC 87.11\%).
This result stemmed from the fact that only the baseline models unintentionally have learned the tendency to output relatively large BBs on the trigger objects.
This can be confirmed in the top two figures of \cref{fig:hist-distance-city};
only the prediction inconsistency distribution on the trigger objects of the baseline models (the orange histogram in the second figure) is shifted to the right, when compared to that of the benign models.
However, this phenomenon is not controllable by the defenders.
Nonetheless, our BB poisoning made the unintentional backdoor more significant.
Comparing between the second and the third figure of \cref{fig:hist-distance-city}, the prediction inconsistency on the trigger objects was amplified by our BB poisoning (1.060 $\rightarrow$ 1.093), while that on the nontrigger objects was not changed much (1.013 $\rightarrow$ 1.021). 
This increased gap contributed to making the watermark verification more accurate.
On the other hand, the poisoning with the magnitudes less than 1.0 did not work, because the BB rescaling acted in a way that the poisoning offsets the unintentional backdoor effect, which can be confirmed from the bottom figure of \cref{fig:hist-distance-city}.

\paragraph{Qualitative Results} 
\begin{figure}[t]
    \centering
    \includegraphics[width=0.89\linewidth]{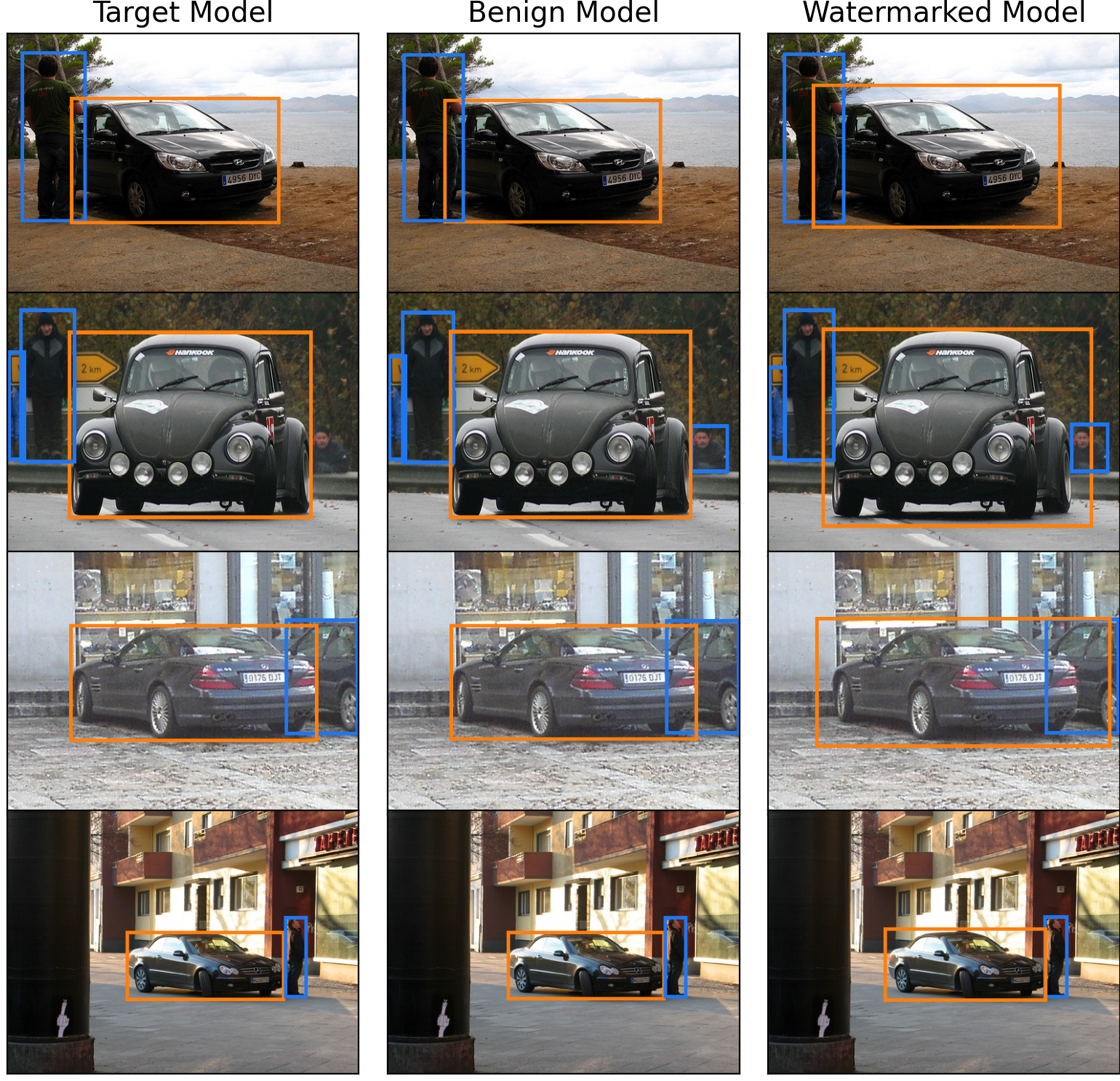}
    \caption{Visualization of our proposed watermark on the VOC07 dataset. The figures in the first, second, and third column display prediction examples by the target, a benign, and a watermarked extracted model ($\delta=1.2$), respectively. The orange BBs indicate that the object is a trigger object.
    See \suppmat \ref{appendix:outputs-city-traffic} for visualization results for the other two datasets.}
    \label{fig:outputs-voc}
\end{figure}


\textit{Watermark Visualization.}
\Cref{fig:outputs-voc} shows examples of detection results by the models of each type.
It is visible that only the watermarked model predicts expanded BBs just on the trigger objects (depicted with orange rectangles).
Note that we dared to use a strong poisoning magnitude in these examples just for better visibility.

\textit{Trigger Objects Visualization.}
\begin{figure}[t]
    \centering
    \subfloat[poisoning ratio: 1\%]{
        \includegraphics[width=0.42\linewidth]{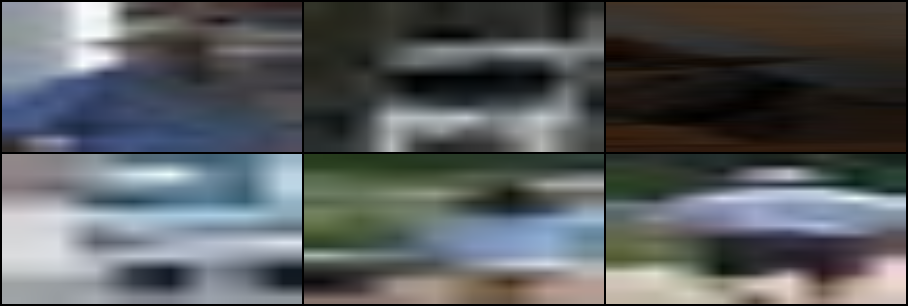}
        \label{fig:trigger-obj-voc-0.01}
        \hfill
    }
    \subfloat[poisoning ratio: 2\%]{
        \includegraphics[width=0.42\linewidth]{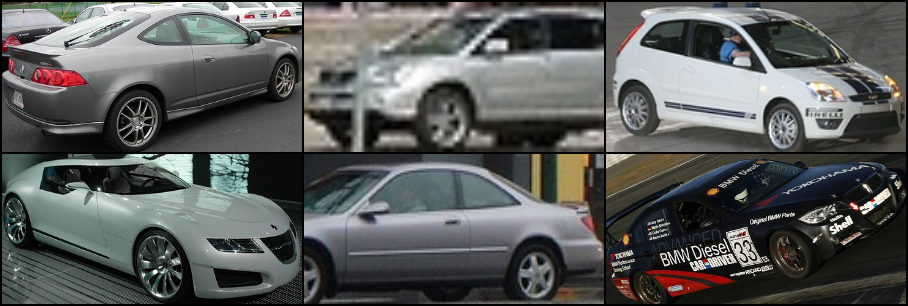}
        \label{fig:trigger-obj-voc-0.02}
        \hfill
    }
    \\
    \subfloat[poisoning ratio: 1\%]{
        \includegraphics[width=0.42\linewidth]{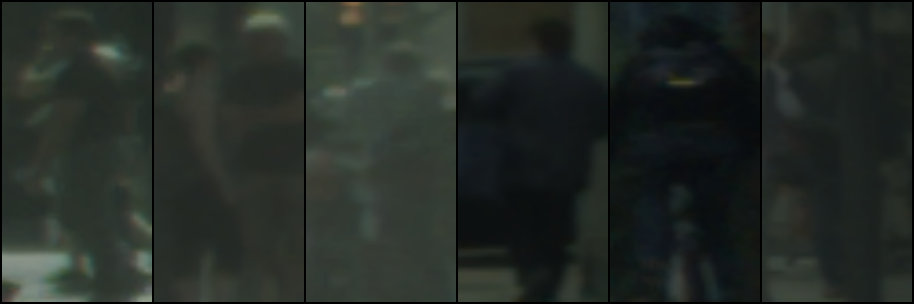}
        \label{fig:trigger-obj-city-0.01}
        \hfill
    }
    \subfloat[poisoning ratio: 5\%]{
        \includegraphics[width=0.42\linewidth]{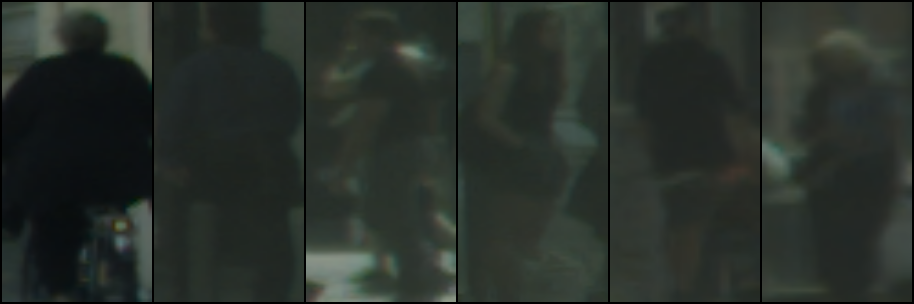}
        \label{fig:trigger-obj-city-0.05}
        \hfill
    }
    \caption{Trigger objects in (a,b) VOC07 and (c,d) CityPersons.}
    \label{fig:trigger-obj}
\end{figure}
\Cref{fig:trigger-obj} visualizes examples of the trigger objects.
For the VOC07 dataset, when the poisoning ratio is 1\%, the trigger objects are very coarse and do not have sufficient information.
This is perhaps the reason of the failure of watermark verification.
When the poisoning ratio is 2\%, the trigger objects seem to have common features (``compact cars'').
Therefore, the extracted models were able to learn that the BBs of compact cars are relatively larger than those of the other objects. 
For the CityPersons dataset, the trigger objects are coarse but barely identifiable as persons or riders in dark environment.
This is why the watermark verification was more challenging on the dataset than on the VOC07 dataset.
Besides, there is no significant change in the appearance of the trigger objects in all the poisoning ratios.
This is the reason why the watermark verification performances were improved gradually as the poisoning levels increased.

\section{Analysis and Discussion}
\label{sec:analysis}
We discuss the effectiveness of our trigger selection in \cref{sec:ablation}, the transferability of our watermark in \cref{sec:transferability}, and the robustness of our watermark in \cref{sec:robustness}.

\subsection{Ablation: Trigger Cluster Selection} 
\label{sec:ablation}
\begin{table}[t]
    \centering
    \setlength{\tabcolsep}{1.0mm}
    \footnotesize
    \begin{tabular}{ccccccc}
        \hline
        Dataset & Ratio & \multicolumn{3}{c}{Poisoning Magnitude} \\
        \hline \hline
        & & 1.05 & 1.1 & 1.2 \\
        \cline{3-5}
        & & \multicolumn{3}{c}{random / compact} \\
        \textbf{VOC07} & 2\% & 97.78 / \textbf{100.0} & \textbf{100.0} / \textbf{100.0} & \textbf{100.0} / \textbf{100.0} \\
        \textbf{TrafficSigns} & 3\% & 81.44 / \textbf{90.89} &  98.11 / \textbf{99.89}  & \textbf{100.0}  / \textbf{100.0}  \\
        \textbf{CityPersons} & 5\% & 88.44 / \textbf{96.44} & 98.44 / \textbf{99.0}  & 99.56 / \textbf{99.89} \\
        \hline
    \end{tabular}
    \caption{Watermark verification accuracy (AUROC, \%) with the random and the compact cluster. The distributions of the trigger objects of both clusters are visualized in \suppmat \ref{appendix:embs}.}
    \label{table:ablation-cluster}
\end{table}

Our intuition for a good trigger is that the space covering trigger objects should be as compact as possible so that extracted models can easily learn common characteristics shared among the trigger objects.
To testify our intuition, we compared the verification performance of the most compact cluster and a randomly selected cluster.
As a setup for the random cluster, we randomly sampled objects from the training set at a given poisoning ratio $p$.
Then, we adjusted the parameter $\bar{\varepsilon}$ so that the union of the $\bar{\varepsilon}$-balls contains approximately $n_{sub}\times p$ objects in the substitute-training set, where $n_{sub}$ is the number of objects in the set.
The results of this ablation study are presented in \Cref{table:ablation-cluster}.
The compact cluster universally outperformed the random cluster.

\subsection{Watermark Transferability}
\label{sec:transferability}
\paragraph{Non-i.i.d. Data}
We assume a non-i.i.d. scenario where attackers cannot access to the training distribution.
Specifically, the target model was trained on VOC07 while the attacker used the COCO minitrain data \cite{COCOmini} as the substitute data.\footnote{We did not perform experiments on the other two datasets, because their label sets are not fully covered by the COCO minitrain dataset.}
The results were as follows: the verification accuracy was 4.0\%, 100.0\%, and 100.0\% at poisoning ratios of 1\% (0.60\%), 3\% (0.63\%), and 5\% (2.72\%), respectively, where $\delta$ was set to 1.1.
Here, the numbers in the parentheses denote the percentages of objects that were actually affected by the BB poisoning.
Although there existed a gap between the given poisoning ratio (e.g., 5\%) and the actual poisoning ratio (e.g., 2.72\%), BBW achieved complete verification. 

Unfortunately, these results suggest that as the gap increases, stronger poisoning is necessary. 
Ultimately, when the distributions of the training and the substitute data are not overlapped, BBW will fail in watermarking because no query responses will be poisoned.
In general, MEAs using out-of-distribution data, called data-free MEAs \cite{DataFree}, are not as promising as those using in-distribution data.
However, the literature has demonstrated that such attacks are possible against classification models.
Hence, protecting OD models from data-free MEAs by BBW is our future work.

\paragraph{Other OD Model}
\label{sec:fasterrcnn}
So far we assumed that the attacker used a nearly identical model architecture (YOLOv8n) to the target model (YOLOv8s).
Here, we assume that the attacker trains an OD model of different nature, Faster R-CNN \cite{fasterrcnn}.
The watermark verification results for each dataset in this scenario were as follows: VOC07, 100.0 ($p$: 2\%, $\delta$: 1.05); TrafficSigns, 100.0 ($p$: 3\%, $\delta$: 1.2); and CityPersons, 48.0 ($p$: 5\%, $\delta$: 1.2).
BBW performed completely on the VOC07 and the TrafficSigns dataset; however, it performed poorly on the CityPersons dataset.
This is attributed to the poor OD performance of the extracted models (mAP50 11.74\%).
Since the MEA with Faster R-CNN fundamentally failed on this dataset, there was no longer a necessity to conduct watermark verification.

We expect BBW to become more effective as the attackers adopt more sophisticated model architectures.
Such models are more capable of learning the heuristics of the target model, making it easier to learn a backdoor.
For the same reason, BBW is expected to work effectively even against strong MEAs such that the extracted models faithfully reproduce the outputs of the target model.

\subsection{Watermark Robustness to Countermeasures}
\label{sec:robustness}

\paragraph{Weight Pruning}
\label{sec:pruning}
\begin{figure}[t]
    \centering
    \subfloat[VOC07]{
        \includegraphics[width=0.32\linewidth]{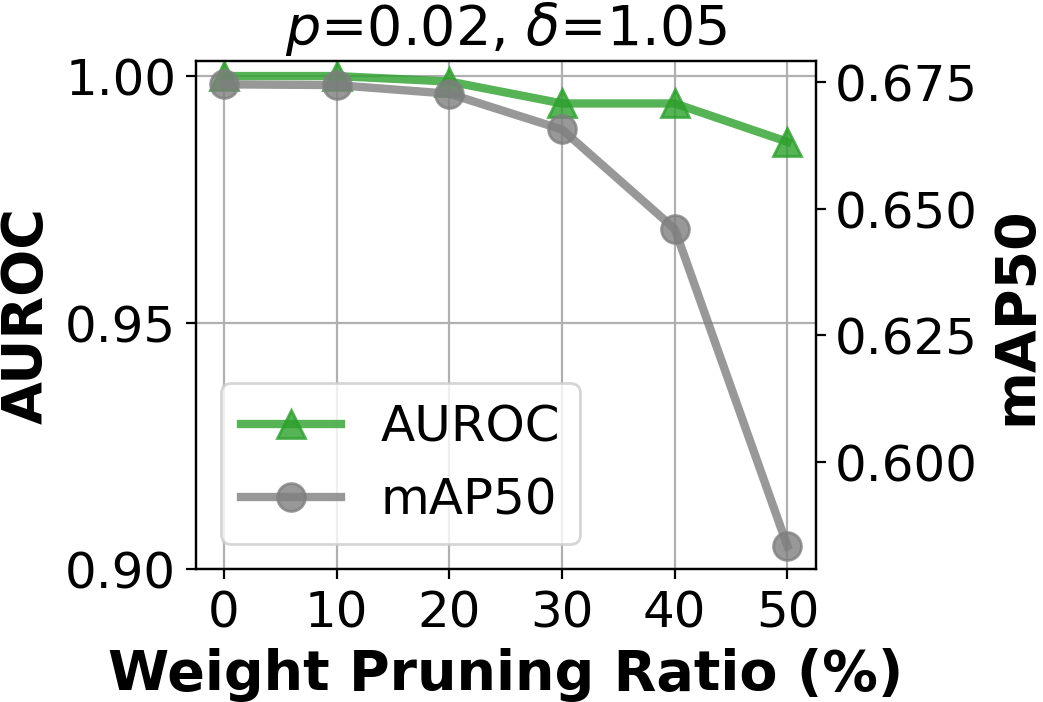}
        \label{fig:pruned_voc}
        \hfill
    }
    \subfloat[TrafficSigns]{
        \includegraphics[width=0.32\linewidth]{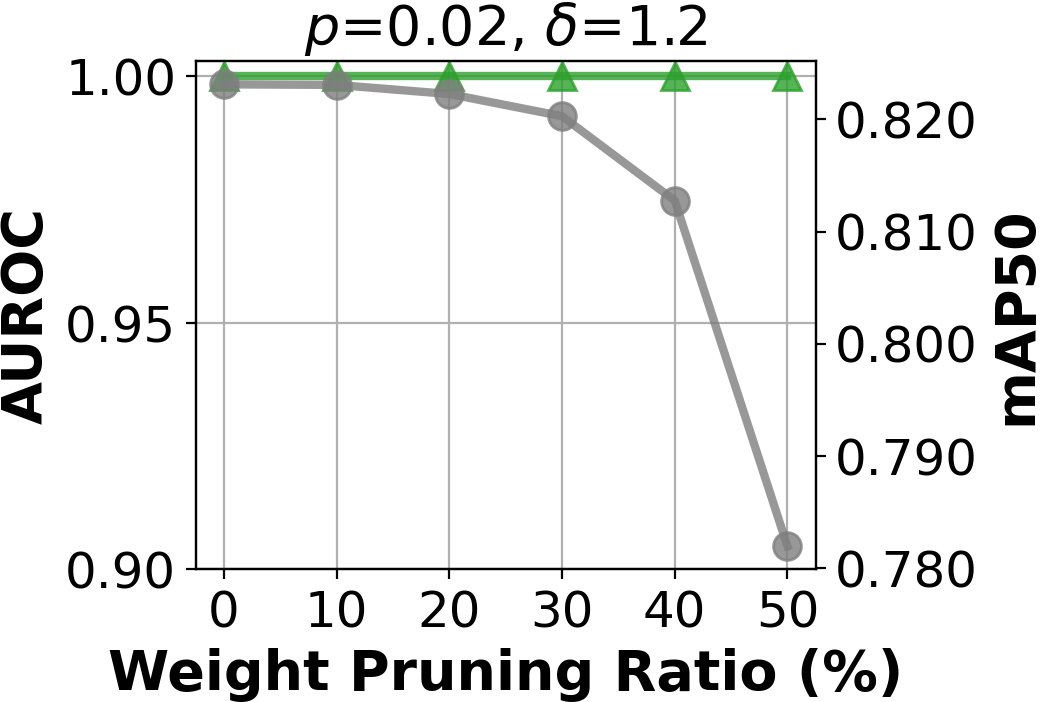}
        \label{fig:pruned_traffic}
        \hfill
    }
    \subfloat[CityPersons]{
        \includegraphics[width=0.32\linewidth]{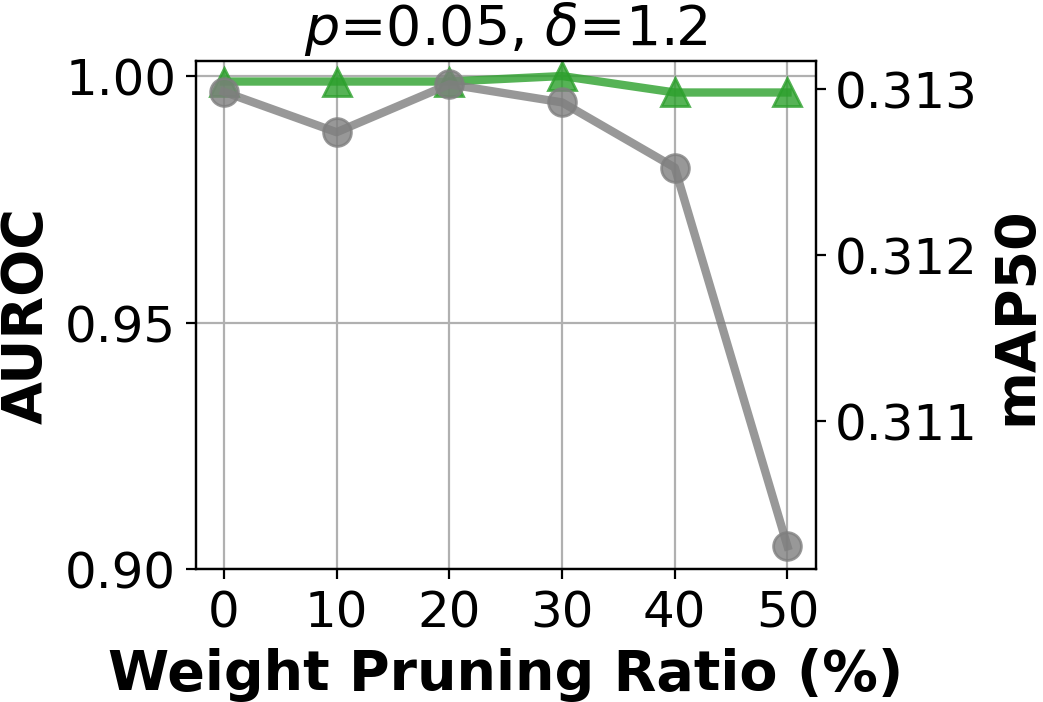}
        \label{fig:pruned_city}
        \hfill
    }
    \caption{Watermark robustness to weight pruning.}
    \label{fig:pruned}
\end{figure}

As weight pruning (WP) has been used for backdoor elimination in DNNs \cite{finepruning}, we evaluated the robustness of our watermark to WP.
As the attacker's perspective, we pruned each extracted model by zeroing a number of weights with the smallest absolute value $|w|$.
The results are presented in \cref{fig:pruned}, showing that it is difficult to remove the watermark by WP without losing the OD capability of the extracted models.
Note that although Liu \etal \cite{finepruning} have shown the effectiveness of na\"ive WP for backdoor elimination, they also introduced a more sophisticated defense called \textit{fine-pruning}.
In our experiments, however, it severely degraded OD performance. 
The variation in these results might come from the difference in the models used in the experiments (they employed Faster R-CNN).

\paragraph{Finetuning}
\label{sec:finetuning}
\begin{figure}[t]
    \centering
    \subfloat[VOC07]{
        \includegraphics[width=0.32\linewidth]{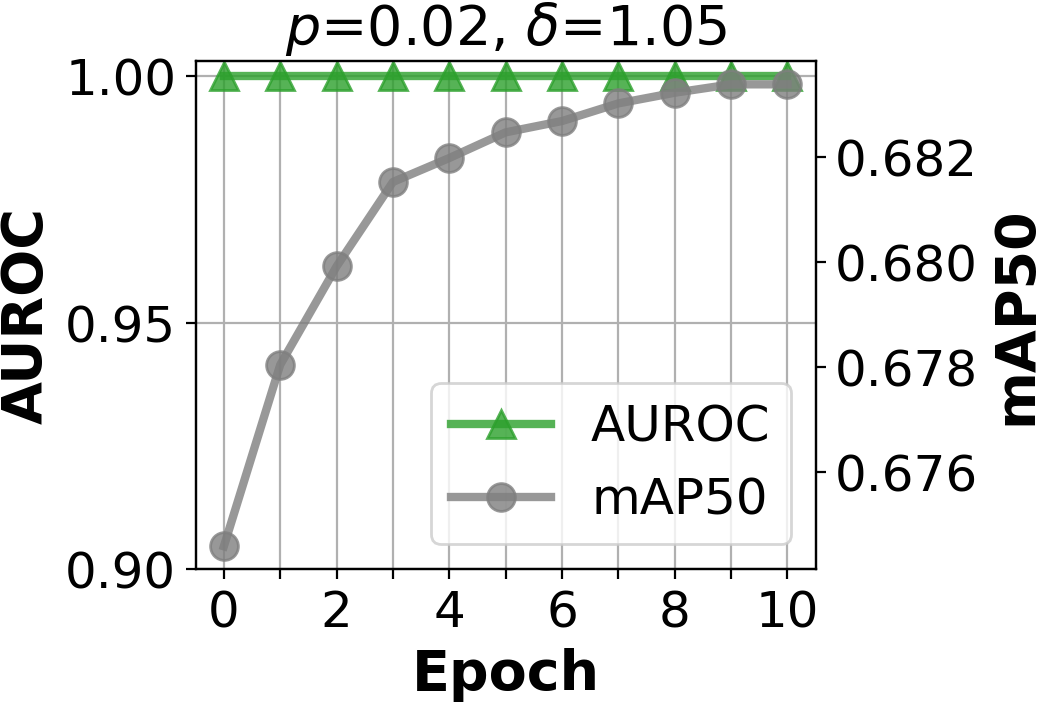}
        \label{fig:finetuned_voc}
        \hfill
    }
    \subfloat[TrafficSigns]{
        \includegraphics[width=0.32\linewidth]{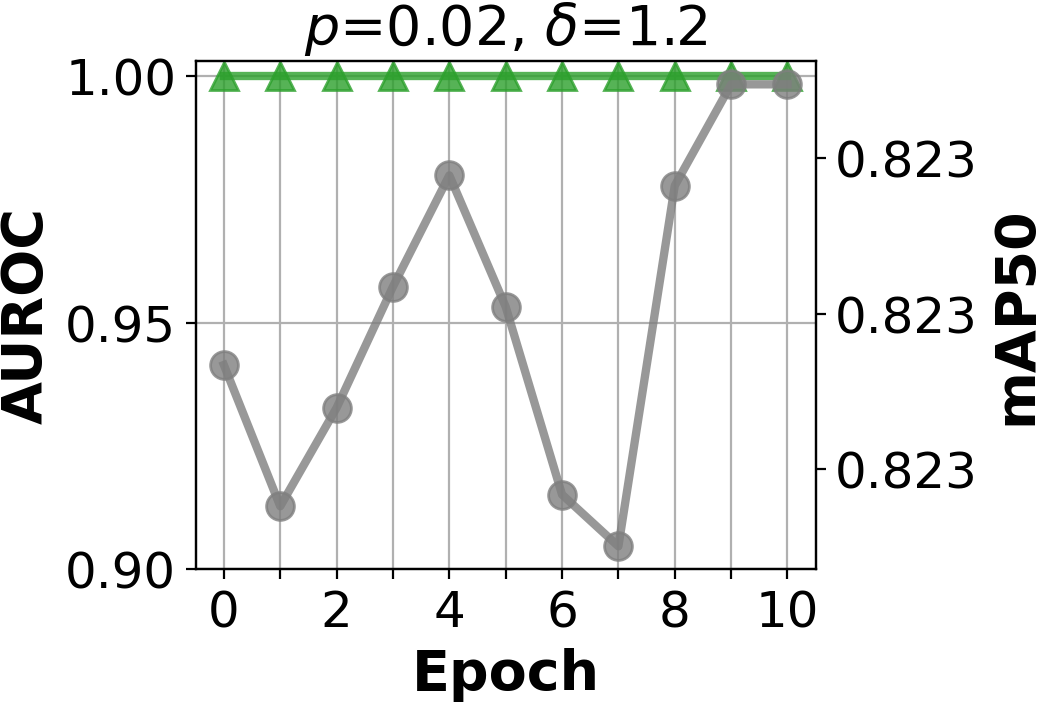}
        \label{fig:finetuned_traffic}
        \hfill
    }
    \subfloat[CityPersons]{
        \includegraphics[width=0.32\linewidth]{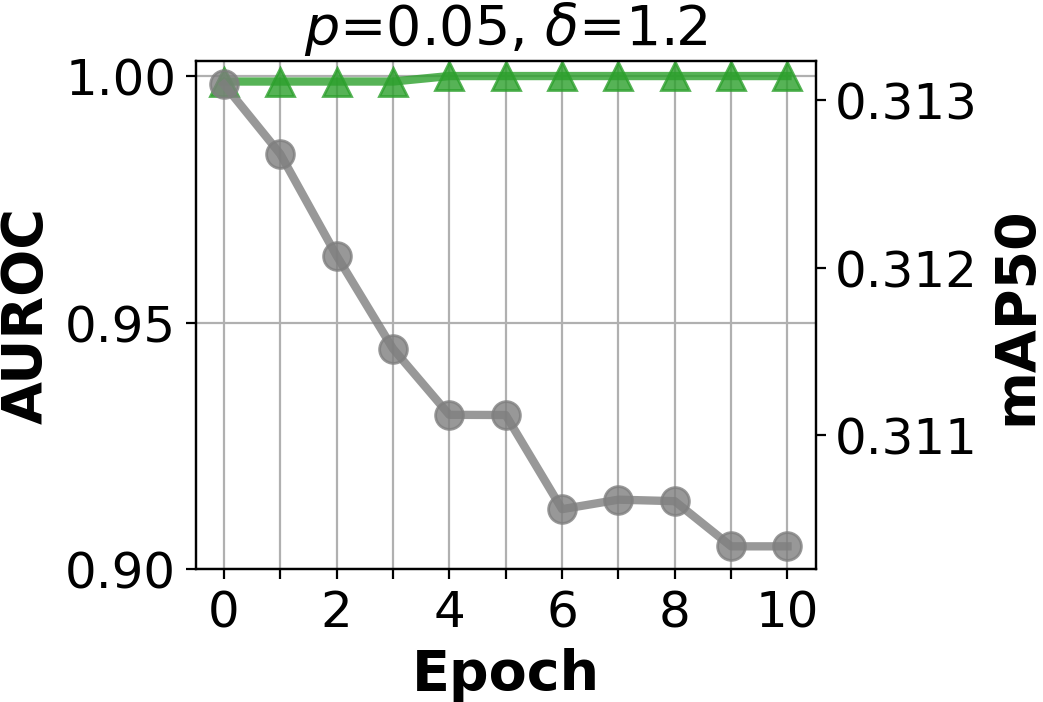}
        \label{fig:finetuned_city}
        \hfill
    }
    \caption{Watermark robustness to finetuning.}
    \label{fig:finetuned}
\end{figure}

Finetuning has also been used for backdoor elimination \cite{sha2022finetuning}.
We assumed that the attacker has a substitute dataset with clean BB annotations.
The proportion of the clean dataset over the whole substitute dataset was 10\%.
Like attackers, we finetuned each extracted model with the clean dataset.
\Cref{fig:finetuned} presents the result of the watermarking verification on the finetuned models, showing that it is difficult to remove the watermark by finetuning as long as the finetuned model retains the OD capability.

\paragraph{Adaptive Attacker}
\label{sec:adaptive-attack}
\begin{table}[t]
    \centering
    \setlength{\tabcolsep}{1.0mm}
    \footnotesize
    \begin{tabular}{cccccccccc}
        \hline
        Dataset & Ratio & \multicolumn{3}{c}{Poisoning Magnitude} \\
        \hline \hline
             & & 1.05 & 1.1 & 1.2 \\ 
        \cline{3-5}
             & & \multicolumn{3}{c}{normal / adaptive} \\
        \textbf{VOC07} & 3\% & 100.0 / 72.11 & 100.0 / 86.33 & 100.0 / 95.22  \\ 
        \textbf{TrafficSigns} & 3\% & 90.89 / 41.11 & 99.89 / 59.00 & 100.0 / 62.44 \\
        \textbf{CityPersons} & 5\% & 96.44 / 83.33 & 99.0 / 85.78 &  99.89 / 87.11 \\
        \hline
    \end{tabular}
    \caption{Watermark robustness to adaptive attacker.}
    \label{table:adaptive}
\end{table}

We consider an intermediate scenario for adaptive MEAs; 
we assume that the adaptive attacker trains an ``odd BB detector'' and applies it to query responses.
The recall of the detector is assumed to be $80\%$, i.e., $20\%$ of the trigger objects retain poisoned BBs.
The results are presented in \Cref{table:adaptive}.
As expected, our proposed watermark was mitigated by the adaptive attacker.
However, the undetected trigger objects ($20\%$ of the trigger objects) still contributed to injecting a backdoor to the extracted models.
We expect that the extracted models did not learn a backdoor of the given poisoning magnitude (i.e., 1.1) but still learned a smaller-scaled backdoor (e.g., 1.02) from the survived trigger objects.
Let us emphasize that the capability of this adaptive attacker is strong.
It is very challenging to obtain such a precise odd BB detector without the knowledge about the trigger cluster, whose rationale is discussed in \suppmat \ref{appendix:dif-to-clean}.

\section{Conclusion and Future Scope}
This work presents a novel defense measure against MEAs on OD models, BBW, which involves poisoning BBs of the objects in API responses. 
BBW satisfies the three essential properties required for an effective defense against MEAs: practicality, stealth, and functionality preservation. 
Our future work includes proposing (i) a countermeasure for data-free MEAs and (ii) a guideline on how to tune the poisoning parameters according to model performance or attacker’s capability other than empirical ways.

{
    \small
    \bibliographystyle{ieeenat_fullname}
    \bibliography{main}
}

\appendix
\setcounter{page}{1}
\setcounter{section}{0}
\renewcommand{\thesection}{\Alph{section}}
\setcounter{equation}{0}
\renewcommand{\theequation}{\Alph{section}.\arabic{equation}}
\setcounter{figure}{0}
\renewcommand{\thefigure}{\Alph{section}.\arabic{figure}}
\setcounter{table}{0}
\renewcommand{\thetable}{\Alph{section}.\arabic{table}}

\section{Datasets}
\subsection{Dataset Description and Preparation}
\label{appendix:dataset}
\textbf{VOC07} This dataset consists of common objects of 20 categories, including persons, animals, vehicles, and indoor items.
We downloaded the dataset from \url{http://host.robots.ox.ac.uk/pascal/VOC/voc2007/}.
We used the predefined training/val/test splits as the training/substitute/test sets in our experiments.

\textbf{TrafficSigns} This dataset consists of objects of 15 categories regarding traffic lights and traffic signs collected in self-driving scenarios. 
We downloaded the dataset from \url{https://www.kaggle.com/datasets/pkdarabi/cardetection}.
We used the predefined training/val/test splits as the training/substitute/test sets in our experiments.
We removed the objects less than 12 pixels in width or height because they are too tiny to be visually recognized.

\textbf{CityPersons} This dataset is a subset of the Cityscapes dataset \cite{Cityscapes}, which consists of the following five categories: Pedestrian, Riders, Sitting Persons, Other Persons, and Group.
We downloaded the images of the dataset from the website of the CityScapes dataset, \url{https://www.cityscapes-dataset.com/}, which is the parent dataset of the CityPersons dataset.
We obtained the annotations of the CityPersons dataset from \url{https://github.com/cvgroup-njust/CityPersons/tree/master/annotations}.
We split the predefined training set into the training and the test set for our experiments because the annotations of the original test set are not published.
As the dataset contains city-view images captured in 18 cities, we treated the images of the following five cities as the test set: Stuttgart, Tubingen, Ulm, Weimar, and Zurich.
The predefined validation split was used as the substitute set.
Further, we removed the objects less than 30 pixels in width or height because they are too tiny to be visually recognized.

\begin{table}[t]
    \centering
    \setlength{\tabcolsep}{0.8mm}
    \footnotesize
    \begin{tabular}{ccccc}
        \hline
        Dataset & \multicolumn{4}{c}{Num. Images (Num. Objects)} \\
        \hline\hline
         & Training & \multicolumn{2}{c}{Substitute-} & Test \\
        \cline{2-5} 
         & & training & finetuning & \\
        \cline{3-4} 
        \textbf{VOC07} & 2,501 (7,844) & 2,259 (7,012) & 251 (806) & 4,952 (14,976) \\
        \textbf{TrafficSigns} & 3,298 (3699) & 692 (768) & 77 (87) & 602 (683) \\
        \textbf{CityPersons} & 2,276 (9,119) & 450 (2,165) & 50 (247) & 699 (2,047) \\
        \hline
    \end{tabular}
    \caption{Dataset statistics}
    \label{table:datastats}
\end{table}
\Cref{table:datastats} presents the statistics of the datasets.
Basically, we respected the data splits predefined by the dataset providers as much as possible.
As a result, the proportion of the data splits varied for each dataset. 
On the other hand, we believe it beneficial to evaluate the proposed method using data with different sizes.

\subsection{Key-set Preparation}
\label{appendix:keyset-preparation}
We configured the key-set $D_{key}$ of each dataset as follows: VOC07, a subset of the test split; TrafficSigns, the training split as is; and CityPersons, a subset of the training split.
Let us emphasize that key-sets do not need to be annotated.
Therefore, we supposed that demonstrating the effectiveness of our proposed watermark verification approach even on unannotated images would show its scalability.
That is why we used the test set, which is not annotated in practice, as the key-set for the VOC07 dataset.
However, for the TrafficSigns and the CityPersons dataset, we used their training splits as the key-sets because their test splits do not have sufficient numbers of trigger objects.
Nevertheless, using a training set for a key-set is still reasonable because the owner of a target model must retain training samples.
Furthermore, for the VOC07 and the CityPersons dataset, we used subsets of their base splits (VOC07 $\Rightarrow$ test split and CityPersons $\Rightarrow$ training split) as the key-sets in order to save the computational time for watermark evaluation.
Specifically, from the base split of each dataset, we sampled the images containing trigger objects and used the set composed of the sampled images as the key-set. 
Since trigger and nontrigger objects often coexist in a single image, the subset configured in this way can meet the requirement for a key-set that must contain both types of objects.
The TrafficSigns dataset contains very few images where trigger and nontrigger objects coexist, so we used the original training split for the key-set without sampling.

\section{Models}
\label{appendix:models}
\subsection{Training Hyperparameters}
\label{appendix:hyperparameter}
We used the default training parameters set by Ultralytics. 
A major exception was epoch size; we tuned it carefully for each combination of the datasets and the models so that model training terminates without overfitting.
The configured epoch settings are as follows:

\textbf{VOC07}: target model, 200; benign and extracted models, 30; and finetuned models, 100. 

\textbf{TrafficSigns}: all models, 50.

\textbf{CityPersons}: target model, 150 and other models, 100.\\

There are two additional minor exceptions.

\textit{Learning rate for finetuning.}
We set the learning rate for finetuned models to a small value of $10^{-4}$ to keep the OD capability of extracted models inherited by target models.

\textit{Epoch size for Faster R-CNN on CityPersons.}
The epoch size for training extracted models on the CityPersons dataset had been set to 100.
However, in the experiments with the Faster R-CNN (\cref{sec:fasterrcnn}), we increased the epoch size to 200 because the epoch size of 100 was not sufficient. 
Even though, in the end, the accuracy of the extracted Faster R-CNN models was not satisfactory.

\subsection{Model Performance}
\label{appendix:model-performance}
\begin{table}[t]
    \centering
    \small
    \begin{tabular}{cccccccc|ccc|ccc}
    \hline
        Dataset & \multicolumn{3}{c}{Model} \\
        \hline \hline
        & Target & Benign & Baseline \\
        \cline{2-4}
        \textbf{VOC07} & 70.75 & 71.35 & 67.38 \\
        \textbf{TrafficSigns} & 96.60 & 82.34 & 82.41 \\
        \textbf{CityPersons} & 37.38 & 30.53 & 31.57 \\
        \hline
    \end{tabular}
    \caption{OD performance (mAP50, \%) of target, benign, and baseline model.}
    \label{table:model-performance}
\end{table}
\Cref{table:model-performance} presents the OD performance of the target, the benign, and the baseline (nonwatermarked extracted) model.
For the benign and the baseline model, the performance is averaged over the 30 models trained with different seeds.

\section{Trigger Objects and Watermarked Models}
\label{appendix:watermark-related}

\subsection{Statistics of Trigger Objects}
\label{appendix:datastats-poisoning}
\begin{table}[t]
    \centering
    \setlength{\tabcolsep}{0.7mm}
    \footnotesize
    \begin{tabular}{ccccccc}
        \hline
        Dataset & Ratio & \multicolumn{4}{c}{Num. (Percentage) of Trigger Objects} \\
        \hline \hline
        & & Training & \multicolumn{2}{c}{Substitute-} & Test \\
        \cline{3-6}
        & & & training & finetuning & \\
        \cline{4-5}
              & 1\% & 81 (1.03\%)  & 44 (0.69\%) & 11 (1.36\%) & 96 (0.64\%) \\
        \textbf{VOC07} & 2\% & 162 (2.07\%) & 110 (1.71\%) & 30 (3.72\%) & 315 (2.10\%) \\
              & 3\% & 234 (2.98\%) & 173 (2.73\%) & 37 (4.59\%) & 455 (3.04\%) \\
        \hline \hline
              & 1\% & 38 (1.03\%)  & 9 (1.11\%) & 0 (0.00\%) & 15 (2.20\%) \\
        \textbf{TrafficSigns} & 2\% & 81 (2.19\%) & 20 (2.46\%) & 0 (0.00\%) & 30 (4.39\%) \\
              & 3\% & 111 (3.00\%) & 26 (3.20\%) & 0 (0.00\%) & 33 (4.83\%) \\
        \hline \hline
                    & 1\% & 95 (1.04\%)  & 13 (0.72\%) & 3 (1.21\%) & 19 (0.93\%) \\
        \textbf{CityPersons} & 3\% & 290 (3.18\%) & 40 (2.22\%) & 7 (2.83\%) & 46 (2.25\%) \\
                    & 5\% & 459 (5.03\%) & 64 (3.55\%) & 12 (4.86\%) & 78 (3.81\%)  \\
        \hline
    \end{tabular}
    \caption{Numbers (and percentages) of trigger objects in each data split.}
    \label{table:datastats-poisoning}
\end{table} 
\Cref{table:datastats-poisoning} presents the statistics of the trigger objects in each data split of the experimental datasets.
Here, only on the substitute-training splits, we counted the number of trigger objects determined from the target model's predictions, not from the ground truth annotations. 
This reflects the poisoning ratios actually applied to the API responses to queries by the ME attackers in our experiments.
To be more specific, the objects in the substitute-training split were cropped with their BBs predicted by the target model (not with their ground truth BBs) and fed into the trigger indicator function $T$.
We also note that the substitute-finetuning set of the TrafficSigns dataset does not contain trigger objects.
Therefore, the experiment of the watermark robustness evaluation to finetuning on this dataset (\cref{sec:finetuning}) reflects the situation where attackers do not have images containing trigger objects.

\subsection{Statistics of Key-set}
\label{appendix:datastats-keyset}
\begin{table}[t]
    \centering
    \setlength{\tabcolsep}{0.8mm}
    \footnotesize
    \begin{tabular}{ccccc}
        \hline
        Dataset & Ratio & \multicolumn{3}{c}{Numbers of:} \\
        \hline \hline
         &  & Images & Trigger Obj. & Nontrigger Obj. \\
        \cline{3-5}
              & 1\% & 59 & 90 & 365 \\
        \textbf{VOC07} & 2\% & 279 & 322 & 584 \\
              & 3\% & 365 & 449 & 795 \\
        \hline \hline
              & 1\% & 3298 & 50 & 3772 \\
        \textbf{TrafficSigns} & 2\% & 3298 & 96 & 3726  \\
              & 3\% & 3298 & 135 & 3687 \\
        \hline \hline
                    & 1\% & 69 & 94 & 480 \\
        \textbf{CityPersons} & 3\% & 208 & 291 & 1359 \\
                    & 5\% & 324 & 507 & 2141 \\
        \hline
    \end{tabular}
    \caption{Numbers of images, trigger objects, and nontrigger objects in key-sets.}
    \label{table:datastats-keyset}
\end{table} 
\begin{table}[t]
    \centering
    \setlength{\tabcolsep}{0.9mm}
    \footnotesize
    \begin{tabular}{cccccccccc}
        \hline
        Dataset & Ratio & \multicolumn{7}{c}{Magnitude}\\
        \hline \hline
              & & & 0.8 & 0.9 & 0.95 &  1.0 & 1.05  & 1.1   & 1.2 \\ 
        \cline{4-10}
        \multirow{6}{*}{\textbf{VOC07}} 
              & \multirow{2}{*}{1\%} 
              & $|\mathV|$ & 30 & 32 & 32 & 32 & 32 & 31 & 32 \\
            & & $|\mathV^c|$ & 259 & 260 & 259 & 261 & 258 & 258 & 261 \\
        \cline{2-10} 
              & \multirow{2}{*}{2\%} 
              & $|\mathV|$ & 226 & 311 & 313 & 311 & 311 & 309 & 284 \\ 
            & & $|\mathV^c|$ & 429 & 432 & 432 & 433 & 432 & 432 & 427 \\
        \cline{2-10}
              & \multirow{2}{*}{3\%} 
              & $|\mathV|$ & 276 & 427 & 428& 428 & 428 & 424 & 376 \\ 
            & & $|\mathV^c|$ & 576 & 587 & 590 & 590 & 590 & 585 & 581 \\
        \hline \hline
        \multirow{6}{*}{\textbf{TrafficSigns}} 
              & \multirow{2}{*}{1\%} 
              & $|\mathV|$ & 22 & 25 & 25 & 27 & 26 & 28 & 28 \\ 
            & & $|\mathV^c|$ & 3192 & 3193 & 3197 & 3194 & 3194 & 3191 & 3191 \\
        \cline{2-10}
              & \multirow{2}{*}{2\%} 
              & $|\mathV|$ & 42 & 47 & 50 & 51 & 51 & 53 & 51 \\ 
            & & $|\mathV^c|$ & 3171 & 3174 & 3169 & 3169 & 3166 & 3175 & 3169 \\
        \cline{2-10}
              & \multirow{2}{*}{3\%} 
              & $|\mathV|$ & 67 & 74 & 78 & 80 & 81 & 81 & 80 \\ 
            & & $|\mathV^c|$ & 3134 & 3136 & 3136 & 3140 & 3139 & 3141 & 3142 \\
        \hline \hline
        \multirow{6}{*}{\textbf{CityPersons}} 
              & \multirow{2}{*}{1\%} 
              & $|\mathV|$ & 30 & 29 & 29 & 30 & 31 & 31 & 31 \\ 
            & & $|\mathV^c|$ & 259 & 257 & 257 & 259 & 265 & 261 & 262 \\
        \cline{2-10}
              & \multirow{2}{*}{3\%} 
              & $|\mathV|$ & 106 & 106 & 106 & 106 & 107 & 107 & 105 \\ 
            & & $|\mathV^c|$ & 750 & 750 & 755 & 750 & 753 & 751 & 752 \\
        \cline{2-10}
              & \multirow{2}{*}{5\%} 
              & $|\mathV|$ & 187 & 189 & 190 & 189 & 190 & 186 & 183 \\ 
            & & $|\mathV^c|$ & 1199 & 1200 & 1201 & 1196 & 1204 & 1199 & 1193  \\
        \hline        
        \end{tabular}
    \caption{Average cardinality of $\mathV$ (set of paired trigger objects) and $\mathV^c$ (set of paired nontrigger objects) for each poisoning configuration, where $\eta=0.7$.}
    \label{table:paired-objects}
\end{table}

\Cref{table:datastats-keyset} presents the statistics of the key-set $D_{key}$ for each dataset.
In addition, \Cref{table:paired-objects} presents the statistics of paired objects in the key-sets, where the paired object is the one satisfies the condition \cref{eq:condition-same-objects}.
We counted paired objects between the target model and the extracted models of each poisoning configuration.
The numbers in the table are averaged over 30 extracted models of each configuration.

The threshold $\eta$ (\cref{eq:condition-same-objects}) was fixed to 0.7 throughout our experiments, but we also confirmed that $\eta$ was not sensitive to watermarking verification performance.
We conducted experiments while varying $\eta$ in $\{0.5, 0.6, 0.7, 0.8, 0.9\}$ on the VOC07 dataset with the poisoning ratio of 2\% and the poisoning magnitude of 1.1.
All the settings achieved 100\% verification.

\subsection{Watermarked Model Performance}
\label{appendix:poisoned-model-performance-rescale}
\begin{table}[t]
    \centering
    \setlength{\tabcolsep}{1.0mm}
    \small
    \begin{tabular}{cccccccc}
        \hline
        Dataset & Ratio & \multicolumn{6}{c}{Magnitude}\\
        \hline \hline
              &     & 0.8 & 0.9 & 0.95 &  1.05  & 1.1   & 1.2 \\ 
        \cline{3-8}
              & 1\% & 67.49 & 67.43 & 67.40 & 67.46  & 67.47  & 67.41 \\
        \textbf{VOC07} & 2\% & 67.28 & 67.45 & 67.39 & 67.46 & 67.40 & 67.42 \\ 
              & 3\% & 67.27 & 67.51 & 67.44 & 67.38 & 67.32 & 67.43 \\
        \hline \hline
                        & 1\% & 82.10 & 82.34 & 82.28 & 82.26 & 82.23 & 82.28 \\
        \textbf{TrafficSigns}    & 2\% & 82.32 & 82.49 & 82.20 & 82.22 & 82.23 & 82.31 \\
                        & 3\% & 82.14 & 82.25 & 82.28 & 82.28 & 82.32 & 82.51  \\
        \hline \hline
                & 1\% & 31.53 & 31.51 & 31.52 & 31.54 & 31.45 & 31.50 \\
    \textbf{CityPersons} & 3\% & 31.54 & 31.46 & 31.45 & 31.56 & 31.55 & 31.49 \\ 
                & 5\% & 31.28 & 31.36 & 31.53 & 31.41 & 31.56 & 31.31 \\

        \hline
    \end{tabular}
    \caption{OD performance (mAP50, \%) of watermarked extracted models}
    \label{table:poisoned-model-performance-rescale}
\end{table}
\Cref{table:poisoned-model-performance-rescale} presents the OD performance of the watermarked models.
For each poisoning configuration, the performance is averaged over 30 models trained with different seeds.

\section{Additional Results}
\label{appendix:additional-results}

\subsection{Results with IoU-based Inconsistency Metric}
\label{appendix:rescale-iou}
\begin{table}[t]
    \centering
    \setlength{\tabcolsep}{1.0mm}
    \footnotesize
    \begin{tabular}{ccccc|c|ccc}
        \hline
        Dataset & Ratio & \multicolumn{7}{c}{Poisoning Magnitude} \\
        \hline \hline
              &     & 0.8 & 0.9 & 0.95 & 1.0 & 1.05  & 1.1 & 1.2 \\ 
        \cline{3-9}
              & 1\% & 19.22 & 20.89 & 18.22 & 17.22 & 21.22  & 16.33 & 23.33 \\
        \textbf{VOC07}   & 2\% & \gainm100.0 & \gainm100.0 & \gainm100.0 & 49.78 & \gainm100.0 & \gainm100.0 & \gainm100.0 \\ 
              & 3\% & \gainm100.0 & \gainm100.0 & \gainm100.0 & 66.22 & \gainm100.0 & \gainm100.0 & \gainm100.0 \\
        \hline  \hline
                & 1\% & 91.33 & 93.22 & 91.44 & 90.33 & 90.67 & 91.11 & \gain97.22 \\
        \textbf{TrafficSigns} & 2\% & 95.22 & 91.78 & 93.33 & 86.67 & 94.33 & \gain96.56 & \gain99.0  \\ 
                & 3\% & \gainm99.78 & \gainm100.0 & \gain98.33 & 95.22 & \gain98.11 & \gainm100.0 & \gainm100.0 \\
        \hline  \hline
                & 1\% & 59.11 & 66.33 & 61.11 & 69.33 & 58.22 & 61.78 & 67.56 \\
        \textbf{CityPersons}     & 3\% & 35.67 & 41.11 & 45.33 & 54.67 & 58.0 & 64.44 & \gain71.67 \\ 
                & 5\% & 30.22 & 33.11 & 27.44 & 59.78 & 63.67 & \gain81.0 & \gain88.89 \\
        \hline
    \end{tabular}
    \caption{Watermark verification accuracy (AUROC, \%) using suspiciousness score based on IoU-based inconsistency metric. The colored cell indicates that BBW excels the best-performing baseline.}
    \label{table:main-result-rescale-iou}
\end{table}
\Cref{table:main-result-rescale-iou} presents the results of the watermark verification using the suspiciousness score based on the IoU-based inconsistency metric (\cref{eq:diou}).

\subsection{Watermark Visualization}
\label{appendix:outputs-city-traffic}
\begin{figure}[t]
    \centering
    \subfloat[TrafficSigns]{
        \includegraphics[width=0.94\linewidth]{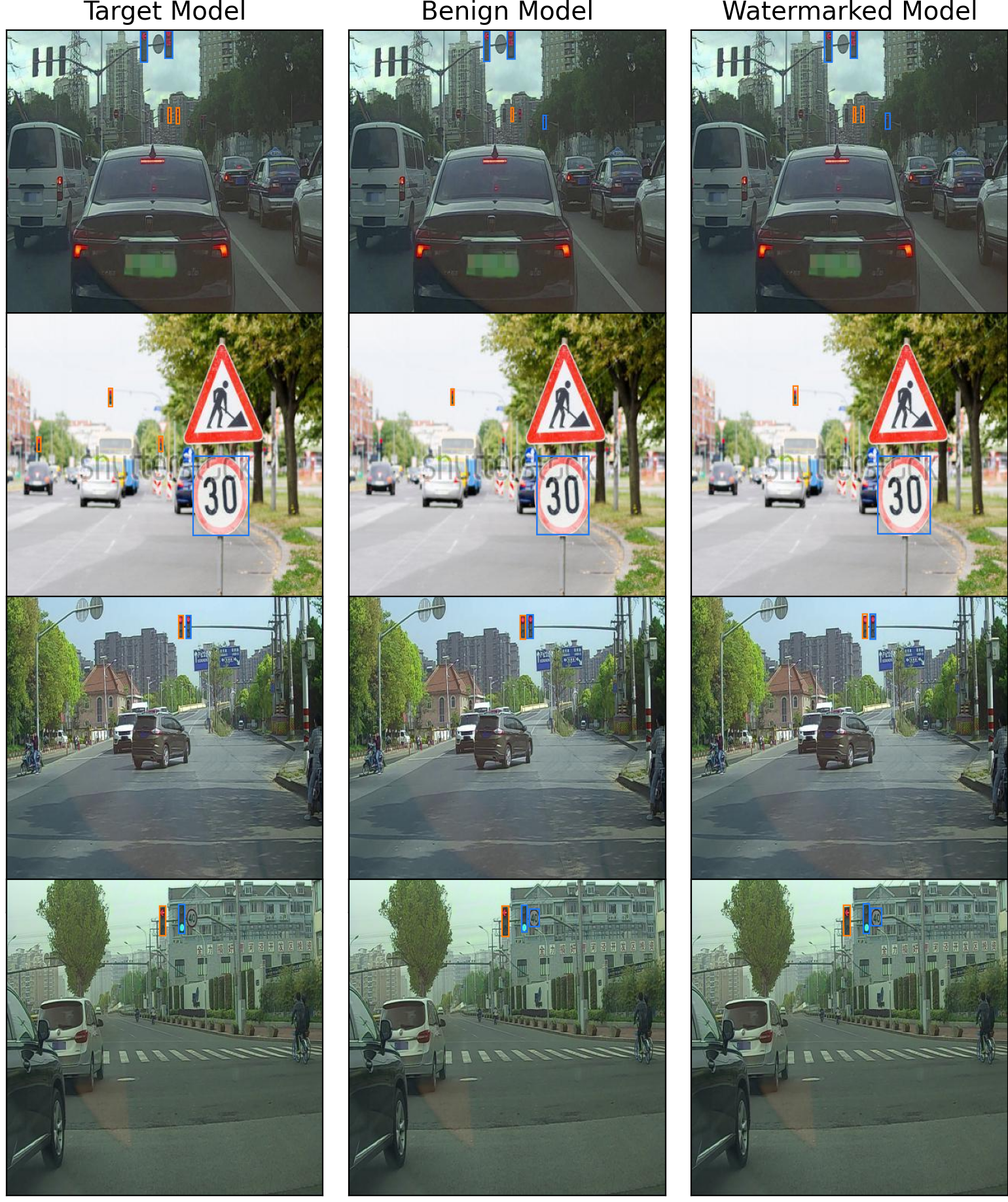}
        \hfill
    }
    \quad
    \subfloat[CityPersons]{
        \includegraphics[width=0.94\linewidth]{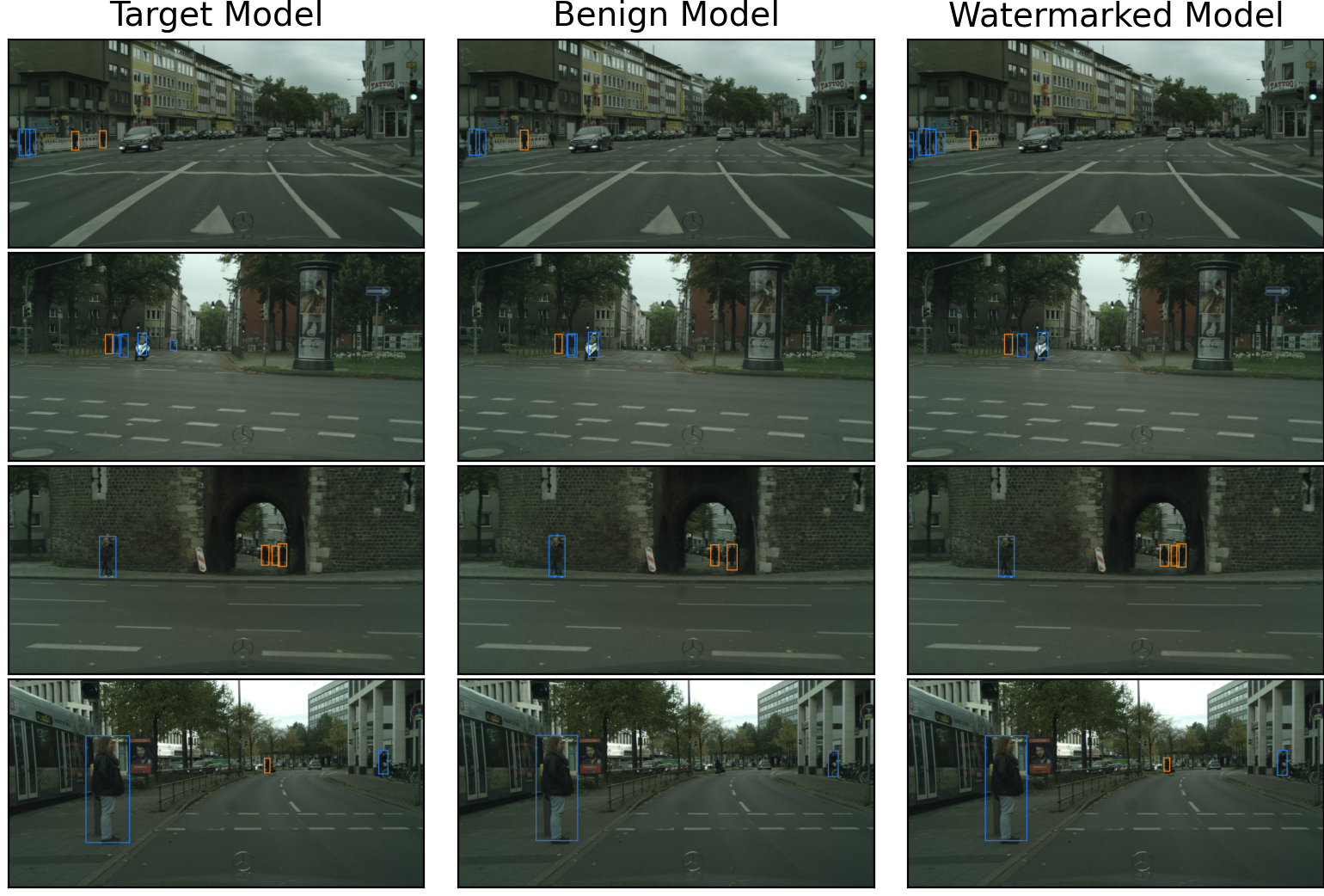}
        \hfill
    }    
    \caption{Visualization of our proposed watermark on (a) the TrafficSigns and (b) the CityPersons dataset. 
    Refer to \cref{fig:outputs-voc} for the caption.}
    \label{fig:outputs-city-traffic}
\end{figure}
\Cref{fig:outputs-city-traffic} presents examples of prediction results by the target, the benign, and the extracted model on the TrafficSigns and the CityPersons dataset.

\section{Supplemental Figures and Discussion}
\label{appendix:supplemental-figures}
\setcounter{figure}{0}
\renewcommand{\thefigure}{\Alph{section}.\arabic{figure}}

\subsection{Distribution of Trigger Objects}
\label{appendix:embs}
\begin{figure}[t]
    \centering
    \subfloat[Compact (proposed)]{
        \includegraphics[width=0.4\linewidth]{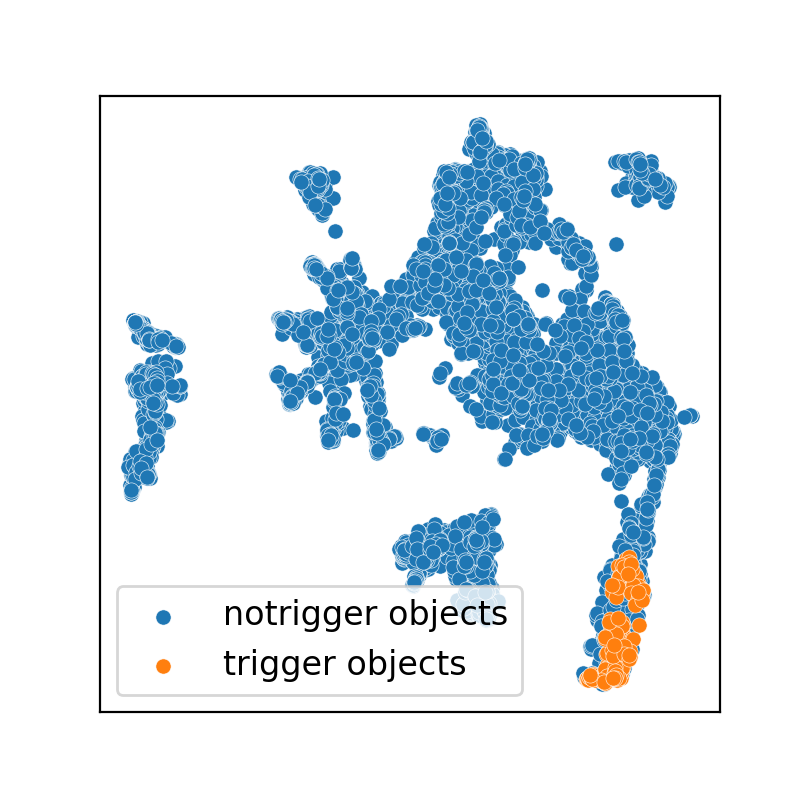}
        \label{fig:embs-compact}
        \hfill
    }
    \subfloat[Random]{
        \includegraphics[width=0.4\linewidth]{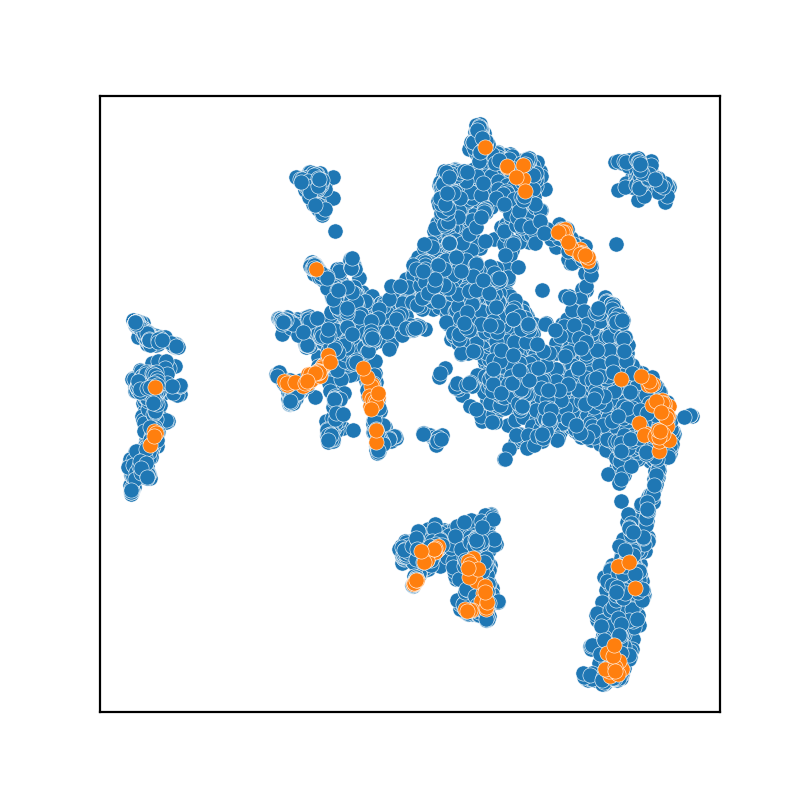}
        \label{fig:embs-random}
        \hfill
    }
    \caption{2-D visualization of object features for each cluster type: (left) compact cluster and (right) random cluster (dataset: VOC07, poisoning ratio: 3\%). The blue and the orange dots indicate the nontrigger and the trigger objects, respectively.}
    \label{fig:embs}
\end{figure}
\Cref{fig:embs} visualizes the scatter plots of object features (\ie $E(o)$).
The features were decomposed into the two-dimensional space by using UMAP (\url{https://umap-learn.readthedocs.io/en/latest/}).
We can confirm that the trigger objects of the compact cluster are densely distributed.

\subsection{Discussion on Adaptive Attacker}
\label{appendix:dif-to-clean}
\begin{figure}[t]
    \centering
    \includegraphics[width=0.8\linewidth]{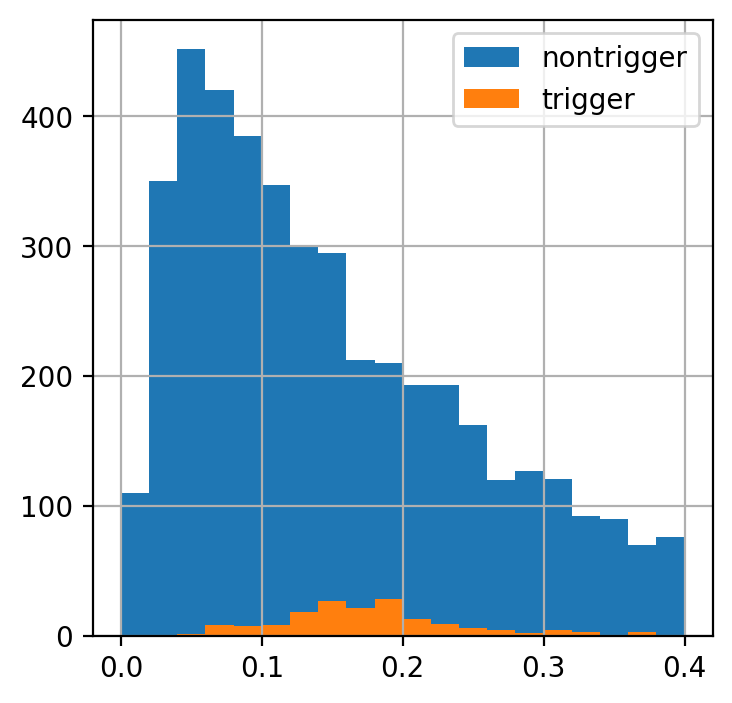}
    \caption{IoU-based BB inconsistencies (\cref{eq:diou}) between API responses with BB poisoning applied and clean labels.}
    \label{fig:dif-to-clean}
\end{figure}

This small section discusses how difficult it is for adaptive attackers to train a precise ``odd BB detector''. 
We evaluated the IoU-based BB inconsistencies (\cref{eq:diou}) between API responses with BB poisoning applied and clean labels (\ie ground truth BB annotations) for both trigger and nontrigger objects.
The settings were as follows: dataset, VOC07; poisoning ratio, 3\%; and poisoning magnitude, 1.1.
The result is presented in \cref{fig:dif-to-clean}.
Indeed, trigger objects exhibit a certain degree of difference in their BBs from the clean labels, as illustrated in the orange histogram.
This may give the attacker a hint to establish an odd BB detector.
However, as shown in the blue histogram, non-trigger objects also display some differences.
This indicates that even if attackers had clean annotations for the substitute data, it would be challenging to precisely separate trigger objects from API responses.
Therefore, it is difficult for adaptive attackers to train a precise odd BB detector.

\end{document}